\documentstyle[emulateapj]{article}

\def \beginfig          {\bigskip}
\def \endfig            {\bigskip}
\def \begineq           {\begin{equation}}
\def \endeq             {\end{equation}}
\def \figcap#1          {{\figcaption{#1}}}
\def \plotone#1         {{\centerline{\epsfxsize=0.95\hsize  {\epsfbox{#1}}}}}
\def \twocolumn         {}

\def \half 	{{1 \over 2}}
\def \d		{\partial}
\def \z		{z}

\def \br	{{\bf r}}
\def \bk	{{\bf k}}
\def \bomega	{{\bf \omega}}
\def \t		{\theta}
\def \bt	{{\bf \theta}}
\def \av#1	{{\langle #1 \rangle}}
\def \max	{{\rm max}}
\def \etal	{{\it et al.\/}}
\def \phys	{{\rm phys}}
\def \diag	{{\rm diag}}
\def \crit	{{\rm crit}}
\def \INT	{{\rm int}}
\def \sinc	{{\rm sinc}}
\def \opt	{{\rm opt}}

\def \rad	{{\rm rad}}
\def \arcmin	{{\rm arcmin}}
\def \zbar	{{\overline \z}}

\def \gsim {\gtrsim}

\begin{document}

\title{Weak Lensing and Cosmology}
\author{Nick Kaiser}
\affil{
Canadian Institute for Theoretical Astrophysics \\
60 St George St., Toronto M5S 3H8 \\
e-mail: {\tt kaiser@cita.utoronto.ca}
}

\begin{abstract}
We explore the dependence of weak lensing phenomena on the background cosmology.
We first generalise the relation between $P_\psi(\omega)$, the angular power
spectrum of the distortion, and the power spectrum of density fluctuations to 
non-flat cosmologies. We then compute $P_\psi$ for various illustrative models.
A useful cosmological discriminator is the growth of $P_\psi$ with source redshift
which is much stronger in low matter density models, and especially in
$\Lambda$-dominated models. With even crude redshift information (say from broad
band colours) it should be possible to constrain the cosmological world model.
The amplitude of $P_\psi(\omega)$ is also quite sensitive to the
cosmology, but requires a reliable external normalisation for the mass
fluctuations. If one normalises to galaxy clustering, with
$M/L$ fixed by small-scale galaxy dynamics, then low density models predict a
much stronger distortion. If, however, one normalises to large-scale bulk-flows,
the predicted distortion for sources at redshifts $Z_s \sim 1-3$ is rather
insensitive to the background cosmology. The signals predicted here can
be detected at a very high level of significance with a photometric survey
covering say 10 square degrees, but sparse sampling is needed to avoid large 
sampling variance and we discuss the factors influencing the design of an
optimum survey. Turning
to weak lensing by clusters we find that for high lens redshifts ($Z_l\simeq1$)
the critical density is substantially reduced in $\Lambda$ models, but that the
ratio of the shear or convergence to the velocity dispersions or X-ray
temperature of clusters is only very weakly dependent on the cosmology.
\end{abstract}


\twocolumn

\section{Introduction}

Weak lensing is the distortion of the shapes and sizes
(and hence fluxes) of distant galaxies from
tidal deflection of light rays by structures
along the line of sight.  
This provides a powerful probe of mass fluctuations
on a wide range of scales from galaxy haloes 
(Valdes \etal, 1984; Brainerd \etal, 1996) 
through clusters 
(Tyson, Valdes \& Wenk, 1990;
Bonnet \etal, 1993;
Bonnet \etal, 1994;
Dahle \etal, 1994;
Fahlman \etal, 1994;
Smail \etal, 1994;
Smail and Dickinson, 1995;
Tyson and Fisher, 1995;
Fort \etal, 1996
Squires \etal, 1996a;
Squires \etal, 1996b;
Dahle \etal, 1996;
Fort \etal, 1996
) 
to supercluster scales 
(Valdes \etal, 1983; Mould \etal, 1994, Villumsen, 1996a).  
As the distortion
depends on the distance to the sources --- being generally larger
for more distant sources --- weak lensing also provides a way
to constrain the redshift distribution for very faint galaxies
(Smail \etal, 1995; Luppino and Kaiser, 1996; Fort \etal, 1996).

The first quantitative predictions for the distortion effect 
were made by Gunn (1967) who was primarily interested in what limits this
placed on the classical cosmological tests. Dyer and Roeder (1974)
estimated the distortion using ``swiss cheese'' models --- with
results quite similar to modern estimates (see below) --- though they
dismissed the effect as too small to have much observational significance.
Webster (1985) computed the increase in ellipticity of distant
objects in models with strong small scale mass inhomogeneity. 
This is related to the weak lensing effects discussed
here, though in fact the broadening of the distribution
of ellipticities only appears at second and higher order
in the gravitational potential $\Phi$, and vanishes in the `weak lensing' limit
considered here.
In weak lensing, one considers the galaxy ellipticity to be a 2-component
vector-like quantity 
$e_\alpha = e\{ \cos 2\varphi, \sin 2\varphi \}$, where $\varphi$ is the
position angle.  The
expectation value of $e_\alpha$ vanishes in the
absence of lensing, and one searches for a 
coherent statistical anisotropy of the
$e_\alpha$ distribution caused by intervening matter.

Quantitative predictions for weak lensing
from modern {\it ab initio\/} models for large-scale structure
were made by Blandford \etal\ (1991)
and Miralda-Escude (1991) who computed the rms
distortion or `polarisation' of distant galaxy shapes and
found that it could be quite large --- on the order of a few
percent in rms shear --- for the then popular theoretical
models such as CDM, and that the rms shear increased as the
$3/2$ power of comoving source distance. These works also
estimated the shear autocorrelation function. In Kaiser (1992,
hereafter K92) this analysis was generalised to include a distribution of
source distances, and extended in a number of other ways:
It was shown how the angular power
spectrum of the shear field was related to the power spectrum
of the density fluctuations in three dimensions; that
shear at observable levels on degree scales was an inevitable
consequence of density fluctuations inferred from large-scale
deviations from Hubble flow, or `bulk-flows' (see Strauss and
Willick, 1995 for a detailed review); that
the (fourier transform of the) projected surface density could
be computed from the shear; and also that one could determine
the M/L for foreground structure by cross correlating the
foreground galaxy density distribution with the shear.
However, this analysis, like those of
Blandford \etal, and Miralda-Escude, was restricted to the
Einstein -- de Sitter cosmological background.
More recently, Villumsen (1996c), and Bar-Kana (1996) have 
discussed the distortion in open
cosmologies. 
Here we will explore in more detail
how the cosmological background affects weak-lensing
observables. In \S\ref{sec:FRWlensing} we give a brief
yet self-contained derivation of the distortion tensor
--- which describes the mapping between angles on the sky at the
observer and position on some distant source plane --- as
a projection of the transverse components of the
tidal field along the line of sight.
In \S\ref{sec:largescalestructure} we consider lensing
by large-scale structure. We first derive the relation
between the angular power spectrum of the distortion and
the power spectrum of the density fluctuations, and
we then consider various illustrative models for $P(k)$
and then discuss the feasibility of these observations
and sampling strategy issues.
In \S\ref{sec:clusterlensing} we consider
lensing by individual clusters.  Our main goal is to elucidate
the dependence of weak lensing phenomena on the background
cosmologies; this will enable us to understand to what extent
our conclusions about the mass distribution and the redshifts
of faint galaxies are cosmology dependent and, to turn the
question around, to see to what extent weak lensing observations,
perhaps combined with other observations, can be used to constrain
the cosmological world model.
In a recent study, Bernardeau \etal\ (1996) have also considered
some aspects of weak lensing considered here, though their work
emphasised more the possibility of measuring higher (than second)
order moments.

\section{Weak Lensing in a FRW Cosmology}
\label{sec:FRWlensing}

We take as the metric for the homogeneous and isotropic FRW background
\begineq
\label{eq:FRWmetric}
ds^2 = g_{\alpha \beta}dr^\alpha dr^\beta 
= - dt^2 + a^2(t) (d\z^2 + \sinh^2 \z d \sigma^2)
\endeq
where $d \sigma^2 \equiv d \theta^2 + \sin^2 \theta d \varphi^2$.
(Here $\z$ measures comoving separation; we will use uppercase
$Z$ to denote redshift).
This is for the open case; for the closed world model
we replace $\sinh \z$ with $\sin \z$.  
The curvature radius $a$ obeys
\begineq
\label{eq:Friedmann}
\left({d a \over dt}\right)^2
= H_0^2 a_0^2
\left(\Omega_m {a_0 \over a} + \Omega_\Lambda {a^2 \over a_0^2}\right)
\pm 1
\endeq
with positive/negative curvature term for open/closed models, and where
$H \equiv (da/dt) / a$ is the expansion rate, a subscript `$0$'
denotes the present value and where
$\Omega_m$, $\Omega_\Lambda$ are understood to be the present values
of the density from matter
and from the cosmological constant in units of the critical value
$\rho_c = 3 H_0^2 / (8 \pi G)$.  Evaluating
(\ref{eq:Friedmann}) at the present we find that the curvature radius,
which we shall use as the scale factor, is related to the current expansion rate by
\begineq
\label{eq:a0}
a_0 = {1 \over H_0 \sqrt{1 - \Omega_0}}
\endeq
where $\Omega_0 \equiv \Omega_m + \Omega_\Lambda$. We will focus on
flat and open models (the former being considered as the limiting case
of the latter), though it is straightforward to generalise the
formulae below to the closed case.

With the addition of small-scale matter inhomogeneity we can take the metric to be
\begineq
\label{eq:FRWmetricplusperturbations}
ds^2 = - (1 + 2 \Phi) dt^2   + (1 - 2 \Phi) a^2(t) (d\z^2 + \sinh^2 \z d \sigma^2)
\endeq
which, on scales much less than the curvature scale,
and in cartesian coordinates, becomes
$ds^2 = a^2 (\eta_{\alpha\beta} - 2 \Phi \delta_{\alpha\beta})dr^\alpha dr^\beta$
where $\eta_{\alpha\beta} = \diag\{-1, 1, 1, 1\}$ 
which is the usual weak-field solution for a source $\delta \rho(\br)$
where $\Phi$ satisfies Poisson's equation
\begineq
\label{eq:poisson}
\nabla^2 \Phi = 4 \pi G \delta \rho
\endeq
and where the Laplacian is taken with respect to proper distance.

Photon trajectories in the spacetime (\ref{eq:FRWmetricplusperturbations})
are solutions of the geodesic equation
\begineq
\label{eq:geodesic}
{d^2 r^\alpha \over d \lambda^2} = - g^{\alpha \beta}
\left(g_{\beta \nu, \mu} - \half g_{\nu\mu, \beta}\right)
{d r^\mu \over d \lambda} {d r^\nu \over d \lambda}
\endeq
To zeroth order in $\Phi$, and for null geodesics,
the time component of (\ref{eq:geodesic}) is 
$d^2 t / d\lambda^2 = -H (dt / d\lambda)^2$ which 
one can solve to obtain the
affine parameter $d \lambda = a dt$.
Let us consider rays confined to a narrow cone around the
polar axis: $\theta \ll 1$, and let $d \sigma^2 = 
d \theta_x^2 + d \theta_y^2$ with 
$\theta_x = \theta \cos \varphi$,
$\theta_y = \theta \sin \varphi$.
The zeroth order solution of the radial
component of (\ref{eq:geodesic}) is $d \z / d \eta = 1$ where
the conformal time is defined as usual by $\eta \equiv \int dt / a$.
The angular components of (\ref{eq:geodesic}), up to first order
in $\Phi$, $d\theta_i/d\eta$ are
\begineq
\label{eq:ddottheta}
{d^2 \theta_i \over d\eta^2} = - 2 {\cosh \z \over \sinh \z} {d \theta_i \over d\eta}
- {2  \over \sinh^2 \z} {\d \Phi \over \d \theta_i}
\endeq
or, in terms of the transverse
comoving displacement of the ray from the polar axis measured in units
of the curvature scale;
$x_i =\theta_i \sinh \z$,
\begineq
\label{eq:ddotx}
\ddot x_i = x_i - 2 \d_i \Phi
\endeq
where $\d_i \equiv \d / \d x_i$ and dot denotes
differentiation wrt conformal lookback time $z$. 
The first term describes the tendency for neighbouring rays to 
diverge due to the hyperbolic geometry, and becomes negligible in the
limit $\Omega_0 \rightarrow 1$,
while the extra forcing term is, as usual, just twice the
transverse gradient of the Newtonian potential.

The general solution of (\ref{eq:ddotx}) is
\begineq
\label{eq:ddotxsolution}
x_i = A_i \sinh \z + B_i \cosh \z 
-2 \int\limits_0^\z d \z' \d_i \Phi(\z') \sinh(\z - \z')
\endeq
which one can readily verify by direct differentiation.
The constants of integration $A$ and $B$ are set by the
boundary conditions. For a
ray which reaches the observer (who we shall place at the
origin of our coordinates) from direction $\theta_{0i}$
these conditions are $B_i = 0$, and $A_i = \theta_{0i}$.

If we consider a pair of neighbouring rays, and assume 
continuity of the
potential $\Phi$, we obtain the
geodesic deviation equation
\begineq
\label{eq:ddotdeltax}
\ddot {\Delta x}_i = {\Delta x}_i - 2 {\Delta x}_j\d_j\d_i \Phi
\endeq
This admits a solution as a perturbative expansion
in $\Phi$ where the $n$-th order term satisfies
\begineq
\label{eq:ddotdeltaxexpansion}
\ddot {\Delta x}_i^{(n)} = {\Delta x}_i^{(n)} - 2 {\Delta x}_j^{(n-1)}\d_j\d_i \Phi
\endeq
Thus, starting with the zeroth order
solution $\Delta x_i^{(0)} = \Delta \theta_i \sinh \z$, one can obtain
the solution for $\Delta x_i^{(1)}$ in the form of 
(\ref{eq:ddotxsolution}), which can then be used to give the
forcing term for the next approximation and so on. 
Here we will restrict attention to the
first order solution:
\begineq
\label{eq:ddotdeltaxsolution1}
{\Delta x}_i = \Delta \theta_i \sinh \z 
- 2 \Delta \theta_j \int\limits_0^\z d \z' 
\sinh(\z') \sinh(\z - \z') \d_j\d_i \Phi
\endeq
This corresponds to evaluating the forcing term using the
zeroth order separation; this being valid either in the `weak-lensing'
approximation where the geodesic deviations are a small
perturbation, or in the
`thin lens' approximation, where the 
ray focusing may become large at great distances but where the 
change in the
separation of the rays as they pass through the lens is small.

Equation (\ref{eq:ddotdeltaxsolution1}) gives the
mapping between angles at the observer
and distance on some distant source plane at $\z_s$:
\begineq
\label{eq:mapping}
{\Delta x}_l(\z_s) = (\delta_{lm} - \psi_{lm}) \sinh \z_s \Delta \theta_m 
\endeq
where we have defined the distortion tensor
\begineq
\label{eq:distortiontensor}
\psi_{lm}(\z_s) = 2 \int\limits_0^{\z_s} d \z  
{\sinh \z \sinh(\z_s - \z) \over \sinh \z_s} \d_l\d_m \Phi
\endeq
in agreement with Bar-Kana (1996).
This is an observable quantity; the traceless parts of $\psi$
causing distortion of shapes of distant galaxies and the
trace causing amplification and hence modulation of the
counts of galaxies.
In reality, we deal with the mean distortion tensor averaged
over $n(\z)$, the distribution of distances to the galaxies
\begineq
\label{eq:psi}
\psi_{lm} = \int d \z_s n(\z_s) \psi_{lm}(\z_s)
= \int d \z g(\z) \d_l\d_m \Phi 
\endeq
where
\begineq
\label{eq:gdefinition}
g(\z) \equiv 2 \sinh \z \int\limits_\z^\infty d \z'
n(\z') {\sinh(\z' - \z) \over \sinh \z'}
\endeq
is a bell-shaped function which peaks at roughly half of the
background source distance,
and where we have normalised $n(\z)$ so $\int d \z n(\z) = 1$.

\section{Large-Scale Structure}
\label{sec:largescalestructure}

We now consider lensing by large-scale structure.  We first
derive an expression for the angular power spectrum of some projected
quantity (be it galaxy counts, image distortion or whatever) and
the corresponding spatial power spectrum. 
This is the fourier space analogue of Limber's equation (K92), 
here generalised to hyperbolic geometries.
We then consider various illustrative models
for $P(k)$ of increasing degrees of realism and then discuss the
feasibility of these observations, sampling strategy and the
prospects for probing large-scale structure {\it via\/} the
amplification rather than shear.

\subsection{Limber's Equation in Fourier Space}
\label{sec:fourierlimber}

A common problem in astronomy is that one observes some
quantity on the sky which is the projection of some three-dimensional
random field or point
process, and one would like to infer the statistical
properties of the latter from the former. An example is
galaxy clustering, where one would like to relate
e.g.~the angular correlation function of the galaxy counts $w_g(\theta)$
to the spatial correlation function $\xi_g(\theta)$.
The solution to this was given by Limber (1954).
Let the projected field be
\begineq
\label{eq:Fdefinition}
F(\bt) = \int d\z q(\z) 
f(\theta_x \sinh \z, \theta_y \sinh \z, \z) 
\endeq
where $f$ is the spatial field written as a function of comoving
coordinates (all in units of the curvature scale) and $q(\z)$ is some
radial weighting function.
The angular two point function of $F$ is
\begineq
\label{eq:xiFdefinition}
\begin{matrix}{
w_F(\Delta \bt) = \av{F(\bt) F(\bt + \Delta \bt)} = 
\int dz \int dz' q(\z) q(\z') \av{f f'} \cr
\simeq \int d \z q^2(\z) \int d\z' 
\xi_f(\Delta \theta_x \sinh \z', \Delta \theta_y \sinh \z', \z'; z)
}\end{matrix}
\endeq
where $\xi_f(\br; z)$ is the spatial two-point function of the
field $f$ at lag $\br$ and conformal lookback time $z$,
and we have assumed
that $q(\z)$ is slowly varying compared to the scale of the density
fluctuations of interest and also that these fluctuations occur on
a scale much smaller than the curvature scale.
This is Limber's (1954) equation, which expresses $w_F$ as an
integral of the spatial two-point function, 
If we fourier transform (\ref{eq:xiFdefinition}) we obtain the
angular power spectrum $P_F(\omega)$.
If we define the transforms
\begineq
\label{eq:transformdefinitions}
\begin{matrix}{
F(\bomega) = \int d^2 \t F(\bt) e^{-i \bomega \cdot \bt} \cr
f(\bk) = \int d^3 r f(\br) e^{-i \bk \cdot \br} 
}\end{matrix}
\endeq
then under the assumption that the fields $F$, $f$ are
statistically homogeneous (or more specifically
that the two point function $\xi_f = \langle f(\br)f(\br')\rangle$
depends only on separation $\br' - \br$) we have
\begineq
\label{eq:fourier2ptfns}
\begin{matrix}{
\av{F(\bomega) F^*(\bomega')} = (2\pi)^2 \delta(\bomega - \bomega') P_F(\omega) \cr
\av{f(\bk) f^*(\bk')} = (2\pi)^3 \delta(\bk - \bk') P_f(\bk)
}\end{matrix}
\endeq
where $P_F(\bomega)$ and $P_f(\bk)$ are the transforms of
$w_F(\theta)$ and 
$\xi_f(\br)$,
so from (\ref{eq:xiFdefinition})
\begineq
\begin{matrix}{
P_F(\omega) = \int d^2 \t w_F(\t) e^{-i \bomega \cdot \bt} = 
\int {d^3 k \over (2\pi)^3} P_f(\bk) \int d\z q^2(\z) \cr
\times \int d^2 \t 
e^{-i(\omega_x - k_x \sinh \z) \t_x}
e^{-i(\omega_y - k_y \sinh \z) \t_y} \int d\z' e^{-i k_z \z'}
}\end{matrix}
\endeq
The angular and $\z'$ integrals here are $\delta$-functions which
pick out the particular spatial frequency $\bk = \{\omega_x / \sinh \z,
\omega_y / \sinh \z, 0 \}$ which contribute to the angular power at
frequency $\bomega = \{\omega_x, \omega_y \}$, and then invoking the
assumed statistical isotropy of $P_f(\bk)$ we have
\begineq
\label{eq:PFdef}
P_F(\omega) = \int d\z {q^2(\z) \over \sinh^2 \z} P_f(\omega / \sinh \z; z)
\endeq
This is the fourier space version of Limber's equation, and is
somewhat simpler than (\ref{eq:xiFdefinition}) as it gives
the angular power spectrum of $F$ as a single integral of the 
spatial power spectrum of $f$,
and provides the generalisation
of (A9) of K92 to hyperbolic geometries. 
As in flat space, it can be thought of
as a convolution in log-frequency space of the three-dimensional
power spectrum of $f$.
Equation (\ref{eq:PFdef}) can be used to relate the
angular power spectrum of galaxy counts to the 3-dimensional
spectrum of galaxy clustering,
in which context $F$ and $f$ would be the density contrast
of galaxies on the sky and in space, and $q(z)$ would be the
normalised distribution of galaxy distances $n(z)$.
Given a specific prediction for $P(k)$, (\ref{eq:PFdef})
enables one to predict $P(\omega)$, or, given suffiently high signal
to noise, one can deconvolve $P(k)$ from $P(\omega)$
(see Baugh and Efstathiou (1994), who used this to
extract the three dimensional power spectrum
of galaxy clustering $P_g(k)$ from the angular power spectrum $P_g(\omega)$
from the APM survey).

What is the advantage of angular power spectrum analysis (PSA) over
the angular auto-correlation function?  
One minor advantage, as we have seen, the
former is somewhat easier to compute from $P(k)$, particularly when
there is evolution of $P(k)$. The real advantage of PSA for
galaxy clustering, however, is that it is easy to
compute the real uncertainty in the power estimates and that the error
matrix for $P(\omega)$ is nearly diagonal; aside from a readily
calculable short range correlation one scales $\delta \omega \sim 1/ \Theta$,
where $\Theta$ is the dimension of the survey, estimates of $P(\omega)$ at
different frequencies are statistically independent.  Neither of these 
pleasant properties hold for correlation analysis. In weak lensing as
we shall now see there is a further advantage in that the observable is
a symmetric $2\times 2$ tensor, and there are various correlation functions
one can form, and making sense of the inter-relation between these is
much simpler in terms of power spectra.

\subsection{Power Spectrum of the Distortion}
\label{sec:powerspectra}

It is now very easy to compute the power spectrum of the
distortion $P_\psi(\omega)$
in terms of the 3-dimensional density field power spectrum $P(k)$,
or equivalently, in terms of $P_\Phi(k)$, the power spectrum
for the potential fluctuations.

The distortion tensor (\ref{eq:psi}) can be written as
the second derivative of a `projected potential':
\begineq
\label{eq:psifromF}
\psi_{lm}(\bt) = 
\d_l \d_m \Phi_p(\bt)
\endeq
where $\d_l$ here and henceforth denotes $\d/\d \theta_l$ and where
\begineq
\label{eq:Phipdefinition}
\Phi_p(\bt) = \int d\z {g(\z) \over \sinh^2 \z}
\Phi(\theta_x \sinh \z, \theta_y \sinh \z, \z) 
\endeq 
which is in the form of (\ref{eq:Fdefinition}) with
$q = g / \sinh^2 \z$.  In fourier space,
differentiation is equivalent to multiplication by $i \bomega$:
$\d_l \Phi_p \rightarrow i \omega_l \Phi_p$
so $\psi_{lm}(\bomega) = - \omega_l \omega_m
\Phi_p(\bomega)$
and therefore the two point function for the distortion is
\begineq
\label{eq:fourierpsi2ptfn}
\av{\psi_{ij}(\bomega) \psi^*_{lm}(\bomega')} =
(2 \pi)^2 \delta(\bomega' - \bomega) 
\hat \omega_i \hat \omega_j \hat \omega_l \hat \omega_m P_\psi(\omega)
\endeq
which nicely factorises into a pure angular term involving the
unit wave vector $\hat \bomega$ --- which, as we will
discuss below, can be
used as a test of the integrity of the data --- and the `distortion
power spectrum':
\begineq
\label{eq:Ppsidef}
P_\psi(\omega) = \omega^{4} \int d \z {g^2(\z) \over \sinh^6 \z} 
P_\Phi(\omega / \sinh\z; z)
\endeq
which is a function only of $|\omega|$.  
Equation (\ref{eq:Ppsidef}) expresses $P_\psi(\omega)$
as a convolution in log-frequency space of the
power spectrum of potential fluctuations $P_\Phi(k)$, or equivalently
to the power spectrum of density fluctuations $P_\delta(k)$, which
is related to $P_\Phi(k)$, through Poisson's equation 
(see (\ref{eq:PpsifromPdelta}) below).
Villumsen (1996c) has obtained an expression similar to (\ref{eq:Ppsidef}), but
finds a different angular dependence.   

\subsection{Models for $P_\Psi(\omega)$}
\label{sec:Pomegamodels}

We now compute $P_\Psi(\omega)$ for various models for the
3-dimensional power spectrum $P_\Phi(k)$.  We first consider
the effect of power concentrated at a single spatial frequency.
We next consider power law models and finally we consider empirical
models which seem to fit most of the data on galaxy clustering
as constructed by Peacock (1996).

\subsubsection{$\delta$-function $P(k)$}

To explore how the background cosmology affects the interpretation
of $P_\psi(\omega)$ let us first compute the angular power spectrum
for a narrow band of power at some frequency with present physical wavenumber
$k_\phys = k_*$, i.e.~$P_\phi(k) = 2 \pi^2  \av{\Phi^2} (a_0 k_*)^{-2}
\delta(k - a_0 k_*)$.  Inserting this is
(\ref{eq:Ppsidef}), and considering the effect on sources at
a single redshift 
($n(\z) \rightarrow \delta(\z - \z_s)$ so 
$g(\z) \rightarrow 2 \sinh \z \sinh(\z_s - \z) / \sinh \z_s$) we find
\begineq
\label{eq:kernel}
P_\psi(\omega) =
8 \pi^2 \av{\Phi^2} \omega
{f^2(\z) \sinh^2(\z_s - \z)  \over \sinh^2 \z_s \cosh \z}
\endeq
where $\z = \sinh^{-1}(\omega / (a_0 k_*))$ and where $P_\psi$ vanishes
for $\omega > a_0 k_* \sinh \z_s$.  Here $f = (a_0 / a) (\delta / \delta_0)$
is the growth factor for the potential, expressed as a function of
lookback time.
Equation (\ref{eq:kernel}) is calculated as follows: Equation
(\ref{eq:Friedmann}) gives $H/H_0$ as a function of $1 + Z = a_0 / a$.
Starting at some very high redshift we compute 
$\eta = \int dt / a = \int da H / a$ to obtain the
conformal lookback time $\z = \eta_0 - \eta$ as a function of $Z$.
We also integrate the equation for the growth of density perturbations:
\begineq
\label{eq:densityperturbationgrowth}
{d^2 \delta \over d t^2} + 2 H {d \delta \over dt} - G \rho_m \delta = 0
\endeq
to obtain $\delta/ \delta_0$ for the growing mode. The result
is shown, for three representative
cosmological models in figure (\ref{fig:ppsiplot1}). In all cases,
the angular power spectrum is a bell-shaped curve, peaking at
roughly half the maximum angular frequency $\omega_\max = a_0 k_* \sinh \z_s$.
Also, the total power increases quite rapidly with increasing
source redshift, this trend being strongest for the $\Lambda$-dominated
models.  

In figure (\ref{fig:ppsiplot2}) we show the total power
\begineq
\label{eq:deltapkvariance}
\begin{matrix}{
\int {d \omega \over 2 \pi} \omega P_\psi(\omega) =
4 \pi \av{\Phi^2} \left({k_*\over H_0}\right)^3 (1 - \Omega_0)^{-3/2} \cr
\times \int\limits_0^{\z_s} d\z f^2(z) 
{\sinh^2 \z \sinh^2 (\z_s - \z) \over \sinh^2 \z_s}
}\end{matrix}
\endeq
and also the
mean (power weighted) angular frequency:
\begineq
\overline \omega \equiv 
{\int d \omega \omega^2 P_\psi(\omega) \over
\int d \omega \omega P_\psi(\omega)} 
\endeq
as a function of source redshift.  At low source redshift the
distortion power spectrum is essentially independent
of the background cosmology, as one might expect, but at high $Z_s$
the low matter density models predict higher distortion.  This is
due in part to the fact that $\Phi$ decreases with time in
these models, and in part to the greater path length back to a given
source redshift.

\beginfig
\plotone{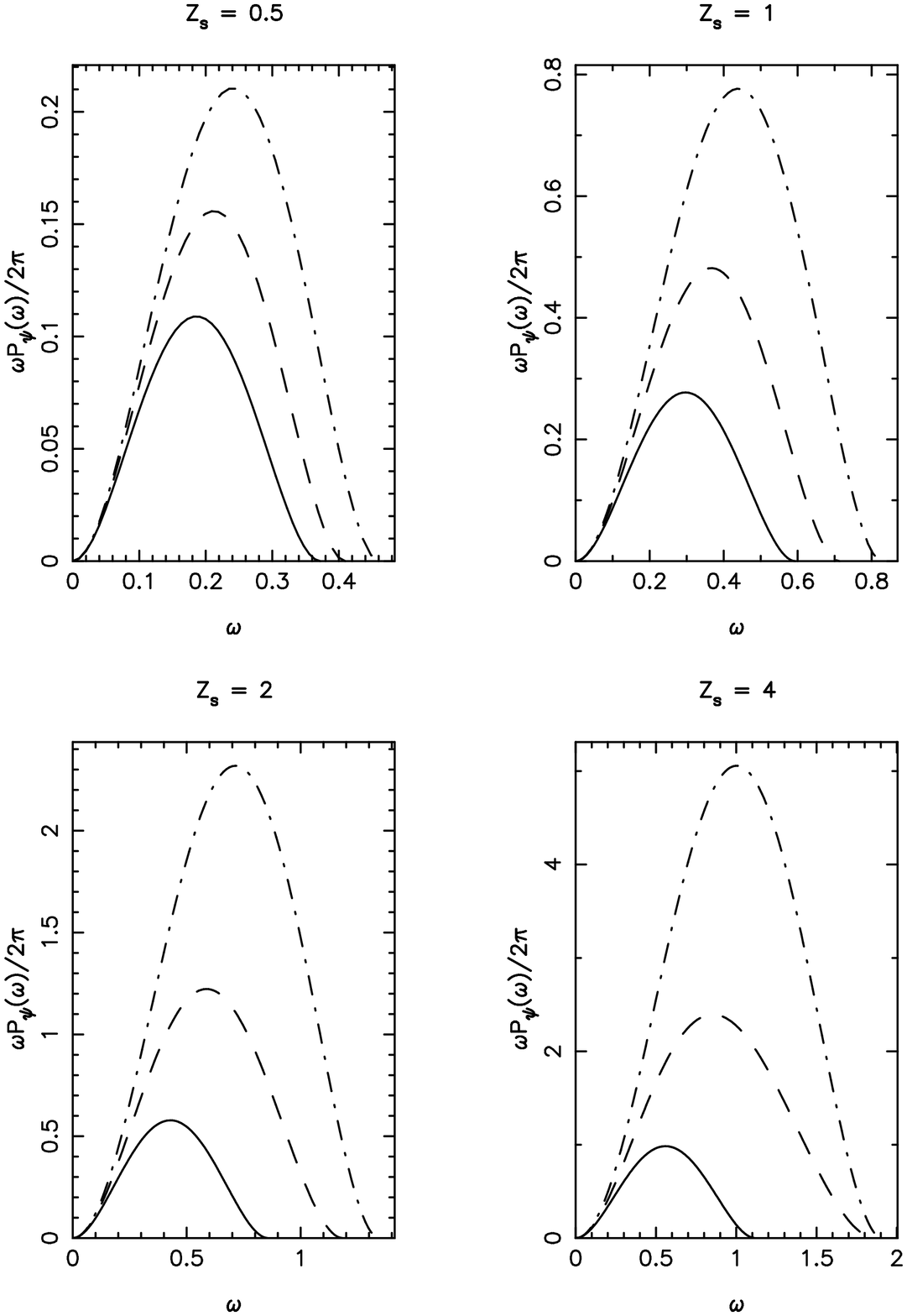}
\figcap{Angular power spectra for a delta-function spatial
power spectrum and for various source redshifts. We have used $\omega
P_\psi(\omega) / 2 \pi$ as ordinate (with linear frequency as the abscissa) 
so that the area under the
curves represents the total power. The solid line is the
EdS model, the dashed line is an open $\Omega_m = 0.2$
model and the dash-dot line is a $\Lambda$ dominated model
also with $\Omega_m = 0.2$.  The numerical values are for
$\av{\Phi^2} = 1$, $k_* = H_0$, so for any other values the
vertical and horizontal scales should be multiplied by
$\av{\Phi^2} (k_* / H_0)^2$ and $k_* / H_0$ respectively.
\label{fig:ppsiplot1}
}
\endfig

\beginfig
\plotone{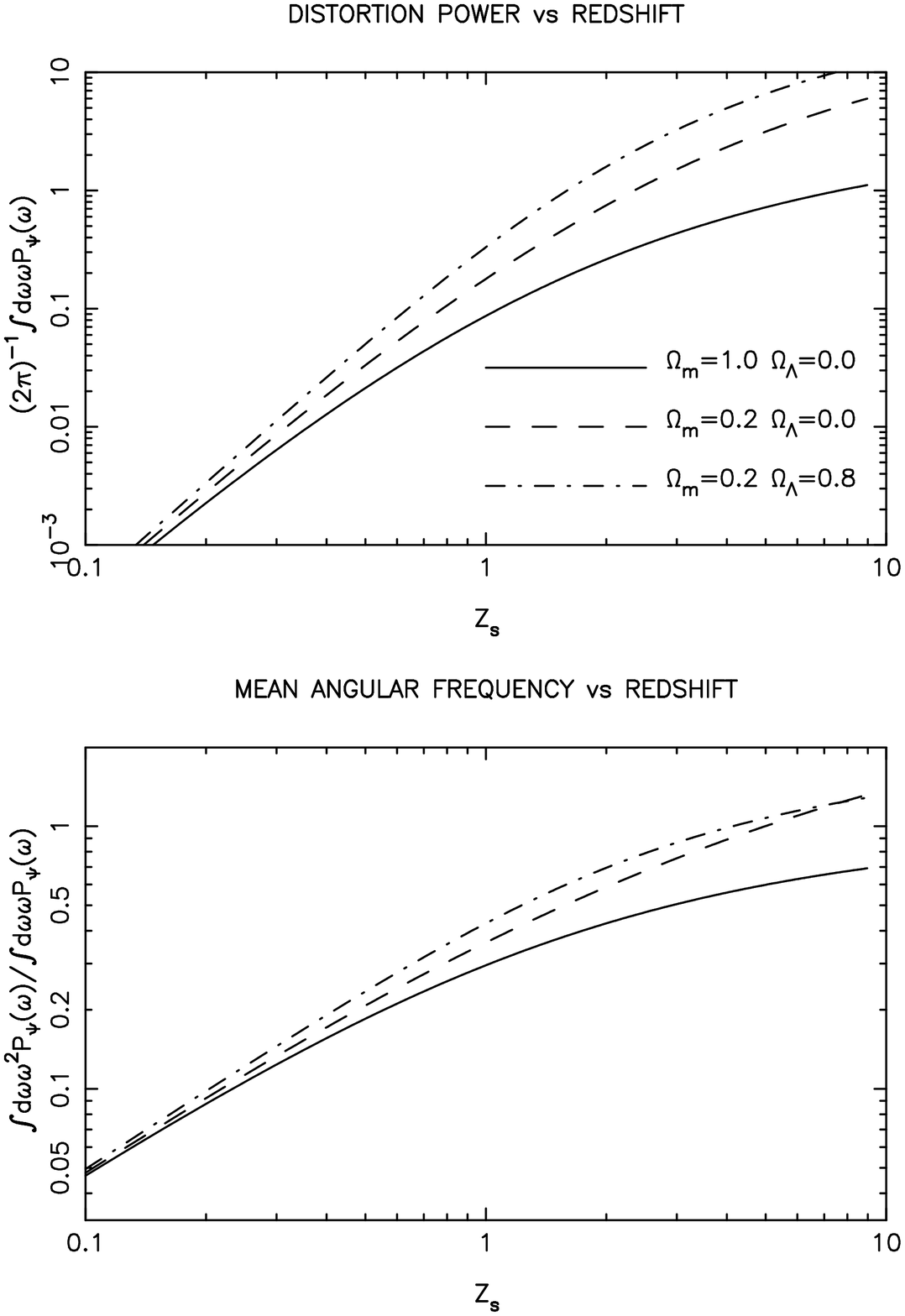}
\figcap{The upper panel panel shows the total distortion power
(for a $\delta$-function potential power spectrum) as a function
of the source redshift. As before, the numerical values assume
$\av{\Phi^2} = 1$, $k_* = H_0$, and one should multiply the vertical 
scale by $\av{\Phi^2} (k_* / H_0)^3$ for other values.
The lower panel shows the power weighted
mean angular frequency for a fiducial physical wavenumber
$k_* = H_0$, which shows that in low matter density models the
power from a given physical scale appears at a somewhat (up to
about a factor two) larger angular frequency.  More striking,
however, is the difference in the total power between the models;
low matter density models, and especially $\Lambda$-models, 
predict much stronger distortion.
This is assuming the same amplitude of 3-D potential fluctuations
for the different cosmologies.  The motivation for this and
alternative normalisations are discussed in the text.
\label{fig:ppsiplot2}}
\endfig

\subsubsection{Power-Law $P(k)$}

These results for a $\delta$-function $P_\Phi(k)$
are most useful to show what spatial
scales we are probing when we measure the angular power at
some frequency. For a realistic $P_\Phi(k)$ we will see a blend
of angular spectra of the form (\ref{eq:kernel}).
As a next step towards realism we now model $P_\Phi(k)$ as
a power law $P_\Phi(k) \propto k^{n - 4}$ (following the usual
convention that the density power spectrum scales as $k^n$). 
To set the
normalisation consistently (corresponding to a given metric fluctuation
variance on a given physical scale), let us take
\begineq
{k^3 P_\Phi(k) \over 2 \pi^2} = 
\av{\Phi^2} _* \left({k_\phys \over  k_*}\right)^{n-1}
= \av{\Phi^2} _* \left({k \over a_0 k_*}\right)^{n-1}
\endeq
so $\av{\Phi^2} _*$ is the contribution to the variance of the potential
per log interval of angular wavenumber at
some fiducial (physical) wavenumber $k_*$, and
which gives, for the distortion power per log interval of wavenumber,
\begineq
\label{eq:powerlawPomega}
\omega^2 P_\psi(\omega) / (2 \pi) = 4 \pi C_n(\z_s) \av{\Phi^2} _* 
\left({k_*\over H_0}\right)^{1-n}
\omega^{2 + n} 
\endeq
where the cosmology dependence is all hidden in the function
\begineq
\label{eq:Cndef}
C_n(\z_s) = (1 - \Omega_0)^{(n-1)/2}
\int\limits_0^{\z_s} d\z f^2(z) 
{\sinh^2(\z_s - \z) \over \sinh^n \z \sinh^2 \z_s}
\endeq
Note that for $n = -2$, which gives equal variance in $\psi$
per log interval of wave-number, (\ref{eq:powerlawPomega}) and
(\ref{eq:Cndef}) are essentially
identical to (\ref{eq:deltapkvariance}). The dependence of
$P_\psi(\omega)$ on source redshift is shown in figure 
(\ref{fig:ppsiplot3}) for various values of the spectral
index $n$.

\beginfig
\plotone{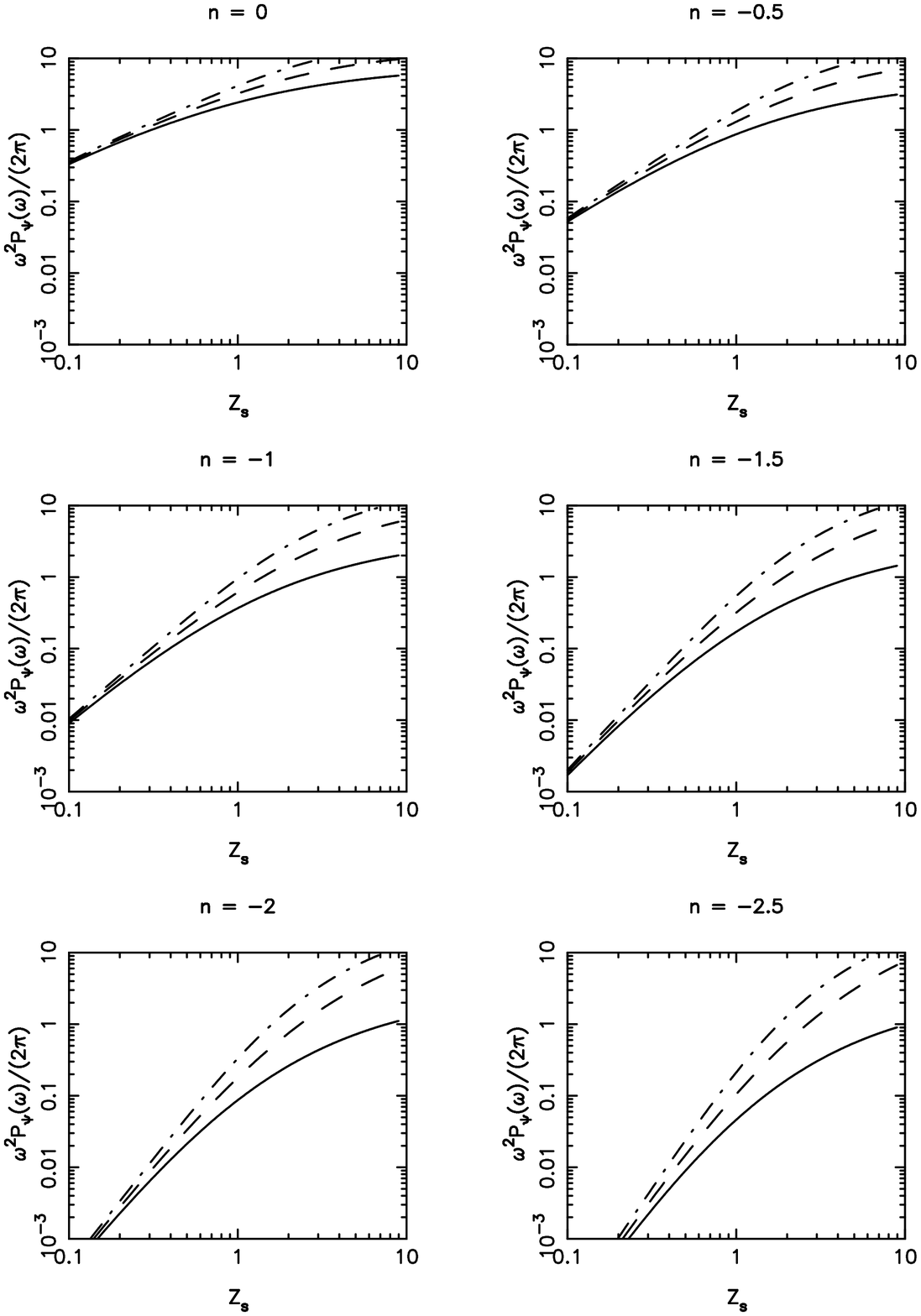}
\figcap{Dependence of the power (per log interval of wavenumber)
on source redshift
for our three example cosmological models now assuming a power law
spectrum of density fluctuations.  The quantity plotted here
is actually $4 \pi C_n$, and should be multiplied by
$\av{\Phi^2} _* (k_*/H_0)^{1-n} w^{2+n}$ to obtain the
real power per log interval of angular wave-number.
\label{fig:ppsiplot3}}
\endfig

\subsubsection{Normalisation}

It is readily apparent that the distortion power is
a strongly increasing function of source redshift in all
cosmologies (though the trend is $n$-dependent).  We also see that
the predicted distortion for high-$Z$ sources is 
sensitive to the cosmology, with low matter density models,
and $\Lambda$-models in particular, giving a much stronger
predicted signal.  This might seem to be at odds with the
conclusions of Villumsen (1996a) and Bernardeau \etal\ (1996),
both of whom find the distortion to be a strongly
{\it increasing\/} function of $\Omega_m$ (Bernardeau \etal\ (1996)
find variance for the top-hat averaged shear approximately
proportional to $\Omega_m^{1.5}$ for instance).  
However, the difference is simply one of
normalisation; we have normalised to a given rms potential
fluctuation whereas Villumsen and Bernardeau \etal\ have implicitly
normalised to a given rms density contrast, 
and had we done this it would have reduced our
predictions by a factor $\Omega_m^2$.  The question of what is the appropriate
normalisation is an interesting one, to which there is as yet
no completely definitive answer.  
Were it the case that all one knew about the
universe was the galaxy correlation function $\xi_g$, then one could reasonably
make a case for normalising to
a given density contrast; implicitly assuming, in the absence of any evidence
to the contrary, that galaxies are unbiased tracers of the mass.
However, there is actually a wealth
of data from dynamical studies of various kinds to suggest that the
appropriate normalisation for the density {\rm contrast} is quite
strongly $\Omega_m$ dependent; in high $\Omega$ models the
galaxy distribution must be biased and it makes sense to fold this
into ones predictions for the rms distortion.

One line of evidence comes from small-scale pairwise velocities;
in the regime where these motions are in equilibrium
these measure the present potential fluctuations directly, and
would therefore lead one to normalise as we have done.  
Thus, if one assumes a scale independent bias,
with $M/L$ tied to small scale `cosmic virial theorem' measurements
(Davis and Peebles, 1983)
then, as shown in figure \ref{fig:ppsiplot3}, 
low $\Omega_m$ models then predict much stronger distortion
at high $Z_s$ (this assumes that one is using
`real-space' estimates of $\xi_g$; if one uses redshift
survey based estimates then one should allow for the
boosting of power due to streaming motions (Kaiser, 1987)). 
Another line of evidence comes from
peculiar velocities on large scale; the so-called
`bulk flows' (e.g.~Strauss and Willick, 1995 and
references therein). The advantage of these observations is that
they directly probe the mass, and give a normalisation to
the mass fluctuations on large-scales where, 
as we shall see, we expect 
weak lensing to be most powerful.  The disadvantage is that
they are very hard to measure reliably and estimates of the
mass power spectrum derived therefrom have very large
`sampling uncertainty' as the observations typically probe only a 
few independent fluctuation volumes.
Here the velocity on a given scale is, crudely
speaking, given by the potential gradient times the age of the
universe, and as the latter is greater in low $\Omega_m$ models
we would infer a lower potential fluctuation amplitude for
a given amplitude for the streaming motions.  A detailed
analysis shows that
this reduces the predicted distortion by roughly a factor
$\Omega_m^{0.4}$ for low $\Omega_m$ models (in amplitude, that is,
corresponding to a factor $\Omega_m^{0.8}$ in power).  
The predicted distortion power 
is shown in figure \ref{fig:ppsiplot3b}  for this normalisation.
Now we find that the greatest distinction between models
appears at very low source redshift, but for sources
at $Z_s \sim 1-3$, and realistic spectral indices,
the predictions are only mildy
dependent on cosmology and hence on $\Omega_m$.
A third line of evidence comes from the abundance of
clusters as a function of velocity dispersion or
X-ray temperature. With e.g.~the Press-Schechter (1974)
model for the mass function, these data can be used to
normalise the amplitude of mass fluctuations
(Cole and Kaiser, 1989). In high $\Omega$ models, and with
the best current data, this gives $\sigma_8 \sim 0.57 \Omega_m^{-0.56}$
(White \etal; 1993) which is very similar to the scaling
from peculiar velocities; however, since low density models
tend to produce power spectra with more large-scale power
(as the horizon size at matter-radiation equality is increased),
the predicted power at large scales in low-$\Omega_m$ models
would not be suppressed
as much as with the bulk-flow normalisation and one would expect to find something
intermediate between the predictions shown in figures
\ref{fig:ppsiplot3}, \ref{fig:ppsiplot3b}.

\beginfig
\plotone{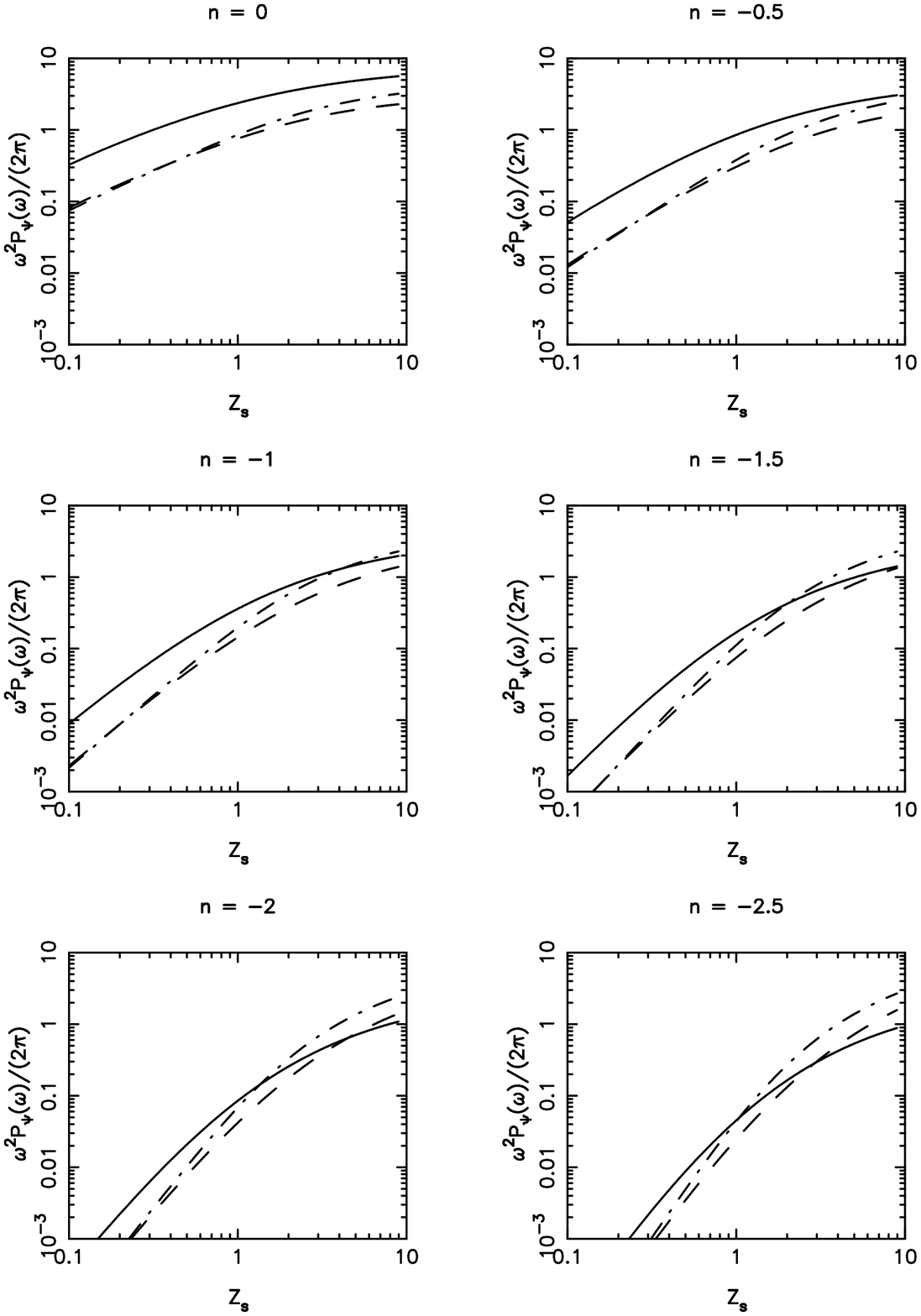}
\figcap{Predicted distortion power versus source redshift
as in figure \ref{fig:ppsiplot3} but now normalised assuming a 
fixed amplitude of bulk-flows.
\label{fig:ppsiplot3b}}
\endfig

There is still considerable uncertainty in the measurements
mentioned above and in their interpretation, 
and so there is considerable slop in
the predicted distortion.
However, we would stress that with any of the
normalisation methods described above,
and for sources at reasonably high redshift $Z_s \sim 1-3$ say,
and for plausible spectral indices $n$ in the range $-1$ to
$-2$ say,
the $\Lambda$ model predictions are at least as high as
in the Einstein - de Sitter model, and in open models
the predicted distortion power is reduced by at most a factor
two and certainly not by a factor $\Omega^{-1.5}$.

\subsubsection{Empirical Models for $P(k)$}

As a final example, we give the distortion power
predicted according to Peacock's (1996) model for the
linear power spectrum, which, for high-$\Omega_m$
incorporates a mild, though plausible level of bias, and
gives a very good fit to most galaxy clustering data.
Poisson's equation (\ref{eq:poisson}) gives 
$P_\Phi(k) = (4 \pi)^2 a_0^4 k^{-4} P_\rho(k) = 
(9/4) \Omega_m^2 k^{-4} (1 - \Omega_0)^{-2} P_\delta(k)$ where
$P_\delta(k)$ is the power spectrum for (mass) density contrast, in terms of
which (\ref{eq:Ppsidef}) becomes
\begineq
\label{eq:PpsifromPdelta0}
P_\psi(\omega) = {9 \pi^2 \Omega_m^2 \over 2 \omega^3 }
\int d\z {g^2(\z) \sinh \z f^2 \Delta^2 \over
(1 - \Omega_0)^2}
\endeq
where $\Delta^2 \equiv k_\phys^3 P^\phys_\delta(k_\phys) / 2 \pi^2$ 
is the  density contrast power per log interval of wavenumber
and should be
evaluated at $k_\phys = \omega H_0 \sqrt{1 - \Omega_0}/ \sinh \z$,
and at conformal lookback time $\z$. 
For sources at a single
redshift this becomes
\begineq
\label{eq:PpsifromPdelta}
P_\psi(\omega) = {18 \pi^2 \Omega_m^2 \over \omega^3}
\int\limits_0^{\z_s}d\z 
{\sinh^3 \z \sinh^2(\z_s - \z) 
f^2 \Delta^2
\over (1 - \Omega_0)^2 \sinh^2 \z_s}
\endeq
This, or more generally (\ref{eq:gdefinition}) and (\ref{eq:PpsifromPdelta0}),
is the most convenient form if one wishes to predict
the distortion from e.g.~COBE normalised {\it ab initio\/}
models.
The results are shown in figure \ref{fig:ppsiplot4} for Peacock's model
for $\Delta^2(k)$, which are quite similar on form to MDM models.
Again, we see that the difference between the different
cosmological models, when normalised realistically, is
quite mild.  The quantity plotted here is $\Delta^2_\psi$, the contribution to
the variance of the trace of $\psi_{lm}$ per log interval, which
is four times the variance in the shear $\gamma$ or the convergence $\kappa$,
so these models are predicting rms shear, convergence $\simeq 1$\%
for $\omega \gsim 100$.  This is similar to the predictions of K92,
though in fact the mass fluctuations assumed here are somewhat lower 
while the adopted redshift
is higher to reflect the growing evidence (both from lensing
and spectroscopy) for a significant
high-redshift component at the relevant magnitude
limits.

The formalism we have developed allows us to compute the
corresponding quantities for any given distribution function
for the background galaxy distances, but, generally speaking, the
results are very similar to the single source plane results
for a single plane at the median distance (see e.g.~K92 for examples).

\beginfig
\plotone{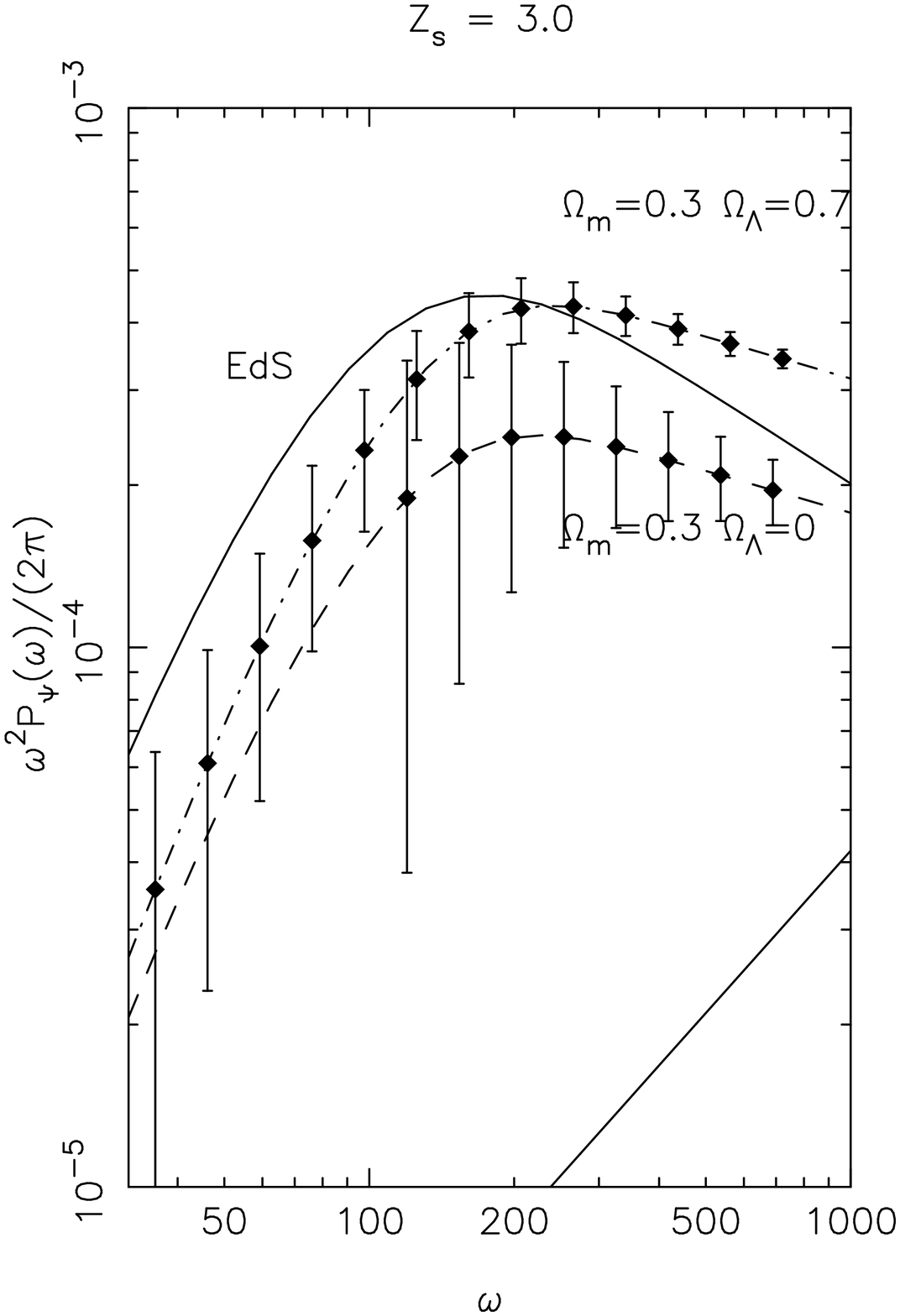}
\figcap{Distortion power according to Peacock's model for $P(k)$.
As before, the solid, dash and dash-dot lines are for
EdS, open and $\Lambda$ models, but here
the low density models have $\Omega_m = 0.3$.  These predictions
are based on a linearised power spectrum.  This should be valid
at large-scales (the strongest distortion here derives from
fluctuations with wavelengths $\sim 100 h^{-1}$Mpc and are
quite accurately linear) but will tend to underestimate the
distortion at small scales where non-linearity acts to
boost the power considerably at late times. The straight
solid line is the expected noise (1-sigma) due to measurement
errors for a 3-degree square survey with realistic
number density and intrinsic ellipticities as described in the text
and for a resolution $d \ln \omega = 0.25$.
The large error bars (arbitrarily attached to the
open model prediction) illustrate the sampling noise for a 
3-degree survey at this resolution.  
These can be reduced considerably by sparse
sampling, as illustrated by the error bars attached to
the $\Lambda$-model which are for a sparse sample of side 9 degrees.
\label{fig:ppsiplot4}}
\endfig

\subsection{Feasibility and Strategy}
\label{sec:feasibility}

We have computed above the power spectrum for the
distortion tensor $\psi_{lm}$.  The quantities we actually
measure are the convergence and shear
\begineq
\label{eq:kappagammadef}
\kappa = \half \psi_{ll} \quad\quad\quad
\gamma_\alpha = \half M_{\alpha lm} \psi_{lm}
\endeq
where
\begineq
\label{eq:Mmatrices}
M_{1lm} = \left[\begin{matrix}{1 & 0 \cr 0 & -1}\end{matrix}\right]
\quad\quad\quad
M_{2lm} = \left[\begin{matrix}{0 & 1 \cr 1 & 0}\end{matrix}\right]
\endeq
The shear $\gamma_\alpha$ is measured from the shapes of galaxies, while
the convergence $\kappa$ can be measured directly from the modulation of the
counts.  The latter tends to be relatively noisy (see Kaiser \etal, 1994),
so we will focus, for the moment, on the shear.

\subsubsection{Shear based $P_\psi(\omega)$}

A procedure for estimating the power spectrum $P_\psi(\omega)$
was outlined in K92. That analysis assumed a
simple square survey geometry.  Here we will generalise this
to more complex survey shapes.  
Let us assume that we have
observations of a set of $N$ galaxies with positions $\{\bt_g\}$ and
that we have measured their shapes to obtain a set of properly
calibrated shear estimates, i.e.~each galaxy provides
an estimate $\hat \gamma_\alpha = \gamma_\alpha(\bt_g)
+ \gamma_\alpha^\INT$ where $\gamma_\alpha^\INT$ is a measure of the
random intrinsic ellipticity of the galaxy plus measurement error. 
The first step is to take the fourier transform of this set of
shear estimates: $\tilde \gamma_\alpha(\omega) = \sum \hat \gamma_\alpha 
\exp(-i \omega \cdot \theta)$, which we can write as
\begineq
\tilde \gamma_\alpha(\omega) = \int d^2 \theta n(\bt) 
\gamma_\alpha(\bt) e^{-i \bomega \cdot \bt} +
\sum \hat \gamma_\alpha^\INT
e^{i \bomega \cdot \bt}
\endeq
where we have introduced $n(\bt) \equiv \sum \delta(\bt - \bt_g)$.
Next we convert this to an estimator of the dimensionless surface density
$\tilde \kappa$ as follows: From (\ref{eq:kappagammadef}) and (\ref{eq:psifromF})
it follows that the gradients of the surface density
and shear are related by $\d_l \kappa = M_{\alpha l m} \d_m \gamma_\alpha$
(Kaiser, 1995) and hence that 
$\nabla^2 \kappa = M_{\alpha l m} \d_l \d_m \gamma_\alpha$, or, in fourier
space $\kappa(\bomega) = M_{\alpha l m} \hat \omega_l \hat \omega_m
\gamma_\alpha(\bomega)$ which suggests the estimator
\begineq
\label{eq:tildekappaestimate}
\tilde \kappa(\bomega) = M_{\alpha l m} \hat \omega_l \hat \omega_m
\tilde \gamma_\alpha(\bomega) = 
c_\alpha(\bomega) \tilde \gamma_\alpha(\bomega)
\endeq
where we have defined $c_\alpha \equiv \{ \cos 2\varphi, \sin 2\varphi \}$
and where $\varphi$ is in turn defined by 
$\omega_i =  \{ \omega \cos \varphi, \omega \sin \varphi \}$.
If one takes the inverse transform of (\ref{eq:tildekappaestimate})
one obtains the surface density estimator of Kaiser
and Squires (1993).  Here we will use (\ref{eq:tildekappaestimate}) 
to obtain an estimate of the
power spectrum $P_\psi$. From 
(\ref{eq:tildekappaestimate}),
(\ref{eq:kappagammadef}), 
(\ref{eq:Mmatrices})
and (\ref{eq:fourierpsi2ptfn}) we find
\begineq
\label{eq:avkappasquared}
\begin{matrix}{
\av{|\tilde \kappa(\bomega)|^2} = N \av{\gamma_1^2} + 
c_\alpha(\bomega) c_\beta(\bomega) \cr
\times \int {d^2 \omega' \over (2 \pi)^2} {P_\psi(\omega') \over 4}
c_\alpha(\bomega') c_\beta(\bomega') |n(\bomega - \bomega')|^2
}\end{matrix}
\endeq
so $\av{|\tilde \kappa|^2} $ does indeed provide an estimate of the power convolved
with $|n(\bomega)|^2$ plus a constant `shot noise' term $N \av{\gamma_1^2} $ where
$\av{\gamma_1^2} $ is the mean square shear
estimate (per component) due to random intrinsic shapes and measurement error.

For a uniformly sampled square survey geometry of side $\Theta$ the 
convolving kernel is
$|n(\bomega)|^2 = N^2 \sinc^2(\omega_x \Theta / 2) \sinc^2(\omega_y \Theta / 2)$
which falls off rapidly for $\omega \gg \Theta$ (and one could
imagine introducing a weight function tapering towards the edge of the
box to apodize the kernel further). For realistic spectra, and for
estimates of the power at $\omega \gg 1 / \Theta$ the convolution
integral in (\ref{eq:avkappasquared}) is then dominated by frequencies very close
to the target frequency (within $\delta \omega \sim 1 / \Theta$), and one
can then to a good approximation remove the relatively slowly varying
factor $P_\psi c_\alpha c_\beta$ from within the integral to obtain
\begineq
\label{eq:avkappasquared2}
\av{|\tilde \kappa(\bomega)|^2}  = 
N\left({\overline n P_\psi(\bomega) \over 4} + \av{\gamma_1^2} \right)
\endeq
where we have invoked Parseval's theorem
and where $\overline n = N / \Theta^2$ is the mean number density of
galaxies per steradian.  
To obtain our final estimate of the power 
we subtract the shot noise term
to obtain $\hat P_\psi(\bomega) = 
4 (|\tilde \kappa|^2 - N \av{\gamma_1^2}) / N \overline n$
and then average these estimates over a shell of frequencies
in $\omega$-space with some width $d \omega = \omega d\ln\omega$.

It is worth noting at this point that if we replace $c_\alpha$ in
(\ref{eq:tildekappaestimate}) by $c_\alpha' = 
\{ \sin 2\varphi, -\cos 2\varphi \}$, or equivalently
apply the 90-degree rotation $\{ \gamma_1, \gamma_2 \} \rightarrow
\{ \gamma_2, -\gamma_1 \}$ to the shear estimates,
then the estimated power
should vanish (aside from statistical noise).  This is a reflection
of the fact that while a general shear field has 2 real degrees of
freedom only one of them is excited by lensing.  Using
$c_\alpha$ in (\ref{eq:tildekappaestimate}) effectively projects out the
active component, while  $c_\alpha'$ projects out the sterile 
component.  This `rotation test' provides a useful
check on the integrity of the data, as most sources of
spurious image polarisation would be expected to excite both
components.  
For further discussion
of this see Kaiser \etal, 1994 and Stebbins, 1996.

To estimate the uncertainty in $\hat P_\psi(\bomega)$ we need to make some
further assumptions about the density fluctuations.  We shall assume  that
the real and imaginary parts of $\tilde \kappa(\omega)$ approximate
a pair of gaussian random fields.  This is certainly appropriate
for density fluctuations arising from inflation, but is also a
very good approximation for many highly non-gaussian (in real space)
processes.  An example is a shot noise process where even for a quite
modest number of shots (say 5 or so), the transform of the
shots becomes quite accurately gaussian
as a consequence of the central limit theorem (see Kaiser and Peacock,
1992, for illustrative examples). The gaussian (transform) approximation
is also valid for density fields which are non-gaussian
due to either non-linear gravitational evolution or biasing. 
Under this assumption we can
calculate the two point function for the `raw' power 
$\av{\tilde \kappa^2(\bomega) \tilde \kappa^2(\bomega + \Delta \bomega)} $
and hence obtain the uncertainty in the shell averaged power.
For a filled square survey, the two point function of the power is,
like $|n(\omega)|^2$, a rather compact
function with width $\delta \omega \sim 1 / \Theta$, 
so estimates of the power at wavenumbers separated by more
than $\delta \omega$ are statistically independent, and computing the
variance in the power averaged over some shell becomes essentially
a counting exercise; one computes $dN _\omega$, 
which is the effective number of independent modes, and then divides the
mean power (signal + shot noise) by $\sqrt{dN_\omega}$
(see Feldman \etal, 1994 for an application of this method to
galaxy clustering in the QDOT redshift survey).

For a simple filled square survey geometry the
number of independent
modes in each shell is just
\begineq
\label{eq:nmodes}
dN_\omega ={\pi \omega^2 d \ln \omega \over (\delta \omega)^2} 
\endeq
where $\delta \omega = 2 \pi / \Theta$ is the fundamental frequency.
Assuming zero signal 
one obtains
a statistical uncertainty in $\Delta_\psi^2 \equiv
\omega^2 P_\psi(\omega) / (2 \pi)$ due to random intrinsic
ellipticities of
\begineq
\label{eq:measurementnoise}
\sigma(\Delta^2_\psi) = \langle (\Delta_\psi^2)^2 \rangle^{1/2} = 
{4 \av{\hat \gamma_1^2} \omega \Theta \over N \sqrt{\pi d\ln \omega}}
\endeq
As noted in K92, for high-$Z$ sources the statistical uncertainty
is very small compared to the expected signal.
For example, with $n \sim 2 \times 10^5$ galaxies per square degree, and
$\av{\gamma_1^2} ^{1/2} \sim 0.40$ (as obtained
from typical cluster lensing studies for integrations of a few
hours on a 4m class telescope) 
and for frequency resolution of say
$d \ln \omega = 1/4$,
or four bins of power per log interval of frequency, we obtain
$\sigma(\Delta^2_\psi) \simeq 1.15 \times 10^{-6} 
(2 \pi  \omega / \Theta)$
which is shown in figure \ref{fig:ppsiplot4} for a survey of side
$\Theta = 3$ degrees.

\subsubsection{Sampling Strategy}

The measurement noise estimate (\ref{eq:measurementnoise}), 
is a tiny ($\sim 1$\%) fraction of the power for our fiducial 3-degree
survey field (assuming Peacock's estimate of the power),
so one would expect, in such a survey, to detect the power
at something like the $\sim 100$-sigma level. This is very nice.
However, it does {\it not\/} imply a similar
precision in determining the true mean Universal power.
For low spatial frequencies the uncertainty
in the ensemble average power will be dominated by the fact that
we only have a small number of independent modes.  
For a 3-degree survey, the fundamental frequency is $\delta \omega
\sim 120$, and the fractional uncertainty in
$P_\psi(\omega)$ at say $2 \delta \omega$ would be around
50\%, rather than 1\%.
This `sampling uncertainty' is shown as the error-bars in figure 
\ref{fig:ppsiplot4} and clearly dominates over the measurement noise.

The situation here is very similar to that in galaxy clustering
studies from redshift surveys 
where even though one might have a sample of many
thousands of galaxies, the number of independent structures on the
largest scales is quite small.  In this situation one can
measure the power in one's sample volume to extremely high
precision, yet the value need not be representative of the
ensemble average power spectrum.  Whether this `sampling uncertainty'
is relevant depends on what one wants to use the data for.
One useful application of redshift surveys is to apply the
cosmic virial theorem and obtain an estimate of the $\Omega$ of
matter clustered like galaxies.  This essentially
involves taking the ratio of the pairwise velocity dispersion
to the galaxy clustering amplitude (Davis and Peebles, 1983).  
Both of these statistics
may have large sampling noise, but their {\sl ratio\/} 
(under the assumption that
there is a universal mass per galaxy) is not affected by this.
Similarly, in weak lensing, one can perform a cross-correlation
between the shear of the faintest galaxies and the surface number density
of somewhat brighter galaxies (chosen so that their $n(z)$
peaks roughly half way to the background galaxies), to obtain an
estimate of the mass to light ratio (K92).  
As with virial analysis, this does not involve the
sample variance, and $M/L$ can be measured to extremely high precision
using a filled survey.
If, however, one's goal is to use $P_\psi(\omega)$ to distinguish between 
e.g.~the three models shown
in figure \ref{fig:ppsiplot4}, then the sampling noise is
clearly a serious handicap.

As with galaxy clustering, sparse sampling (Kaiser, 1986)
could be quite helpful here.  Imagine one were to observe a similar number
of galaxies, but with $N_f$ sparsely spaced fields of size $\Theta_f$
scattered over a much larger square (of side denoted by 
$\Theta$ as before). The function $|n(\bomega)|^2$, which should be
thought of as a `instrumental point spread function' through which we
measure the power will now be much narrower (by a factor $\sim \sqrt{f}$,
where $f$ is the areal filling factor of ones survey).  This has two
benefits; first, the fundamental frequency would decrease,
so one would be able to probe beyond the peak in the spectrum (clearly
an interesting and cosmology dependent attribute). Second, in each region
of frequency space one will obtain $1/f$ times as many independent
estimates of the power, so the sampling uncertainty decreases
by a factor $1 / \sqrt{f}$, which can be a considerable gain.

There is a price, however, for this increased resolution and precision,
which is `aliasing'.  For a sparse survey, 
the convolving kernel $|n(\omega)|^2$ 
will now have side-lobes in addition to the central peak which
extend to relatively high frequencies $\omega_f \sim 2 \pi / \Theta_f$, 
and our power estimator
$|\tilde \kappa(\omega)|^2$ will in general
contain some contribution from $\bomega' \ne \bomega$.
The structure of the side-lobes depends on how one lays out
ones fields.  
If this is done in a random or semi-random manner (perhaps by choosing
a set of fields which avoid few bright foreground objects) then the
side lobes look like (the square of) a random gaussian field
smoothed on a scale $\delta \omega$ and the mean
strength of the side-lobes is 
suppressed by a factor $\sim 1/N_f$
as compared to the central peak.  
If the fields are laid out on a grid then the sidelobes are as strong
as the central lobe, but are spaced on a grid of spacing $\sqrt{N_f} \delta \omega$
and cover only a fraction $f$ of the frequency plane.
Some examples are shown in figure (\ref{fig:surveys}).

There are two aspects to the `aliasing problem'; 
the first is mixing of power from frequencies similar to the target
frequency: $\bomega' \sim \bomega$. For a random or grid-like survey
this is small provided
$N_f \gg \omega^2  \Theta^2$, and this condition says
that one must sample several fields
per wavelength of interest.  If the fields are laid out in a line,
a substantial mixing of power is unavoidable (as is the case
for pencil beam or 2-dimensional redshift surveys).  This is not
absolutely disastrous, as one can always convolve one's theoretical
predictions, but it seems an unwanted and unnecessary complication,
and we would discourage this.
Also, in order to apply the
`rotation test' described above it is necessary one has proper
2-dimensional sampling and that the above condition be satisfied.

\beginfig
\plotone{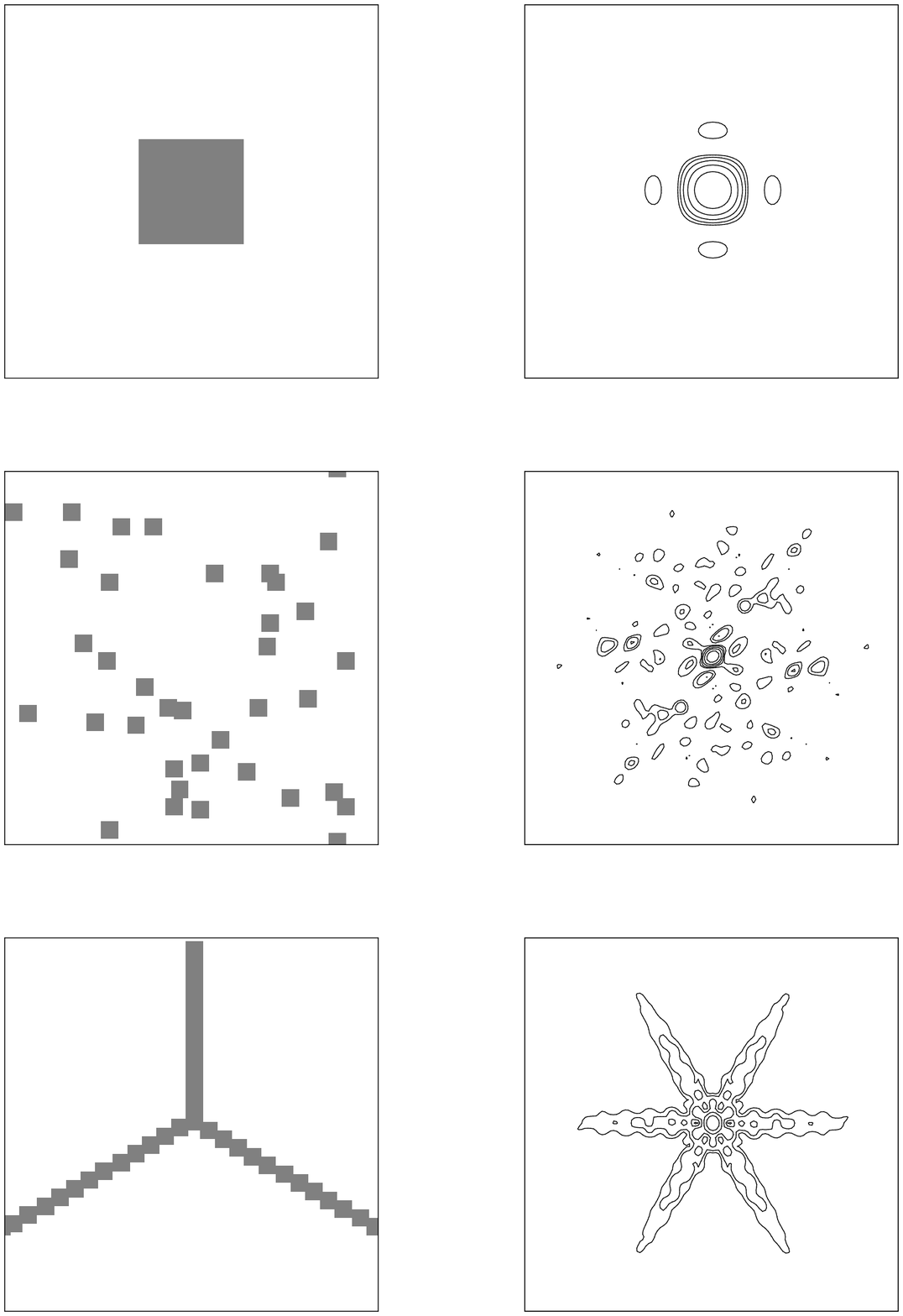}
\figcap{The panels on the left illustrate three possible ways one might
lay down fields for a weak lensing survey.  In each case there are
$N_f = 36$ fields, and, assuming a field size of $0.5$ degrees the box side
would be about 10 degrees.  The top panel shows a filled survey
(3 degrees on a side), the middle panel shows a sparse survey
with randomly placed fields while the bottom panel shows 
a `VLA' style layout.  The panels on the right show the
corresponding `point spread functions' for power spectrum estimation.
The contours are logarithmic at $2^{-n}$ times the peak value.  The filled survey
has a wide central lobe, and hence very poor precision as the sampling
of the power spectrum is very coarse (as discussed in the text, the
variance in the power is inversely proportional to the
area of the central lobe), but the aliasing of high frequencies
is very small.  The random sparse survey has a very tight central
lobe, and would yield a factor $\sim 3$ improvement in precision
over the filled survey. There is also rather
good rejection of neighbouring frequencies close to the target frequency, but one
can see the sidelobes (the typical height of which is suppressed
by a factor $1/N_f$ relative to the central lobe) which extend to
frequencies of order the inverse field size.  For the models discussed
here the power aliased through these sidelobes would be very small.
The `VLA' style survey has very poor behaviour and combines the
worst features of the two strategies above; the central lobe is
not much narrower than the filled case, which yields large sampling
variance, while the side lobes are at least as strong as for the random
sparse survey.  Moreover, there is poor rejection of frequencies close
to the target frequency, so the measured power spectrum at low frequencies
will be quite distorted and one would need to apply deconvolution.
\label{fig:surveys}}
\endfig

The second aspect is aliasing of power
from higher frequencies $\omega' \gg \omega$.  
This is somewhat more model dependent
as it depends on how the power varies with frequency. 
One finds from (\ref{eq:avkappasquared}) that the aliased
power will be small compared to the power one is measuring
from the target frequency, provided $N_f \gg 
\omega^2 \Theta^2 \Delta_\psi^2(\omega') / \Delta_\psi^2(\omega)$.
This again is physically reasonable; $N_f / (\omega^2 \Theta^2)$ is the
number of fields per square target wavelength, so
when this condition is only marginally
satisfied the small scale power induces a `root-$N$' contribution
in the low-$\omega$ $\kappa$ estimate which is equal to the intrinsic
low frequency $\kappa$ fluctuations.
Thus, if $\Delta_\psi^2(\omega)$ increases strongly towards small
scales then one would want to increase the sampling density. 
However, if the models shown in figure \ref{fig:ppsiplot4} are a 
realistic guide
$\Delta_\psi^2$ is actually falling with increasing frequency,
so aliasing from small scales would not be a serious problem, and
a rather sparse sampling would pay great dividends with
little or no cost.  It may be that this is overly optimistic, since,
as mentioned above, non-linearity may boost
the power at small scales at late times and cause the predictions
shown in figure \ref{fig:ppsiplot4} to underestimate the
lensing power. Nonetheless,  it should still possible to choose
$N_f$ so as to make the aliased power small.  Assuming these conditions
have been met, the power estimate is
\begineq
\label{eq:Ppsisparse}
\hat P_\psi(\bomega) = {4 \Theta^2 \over N^2}
\left(|\tilde \kappa|^2 - N \av{\gamma_1^2} \right)
\endeq
and the uncertainty in the distortion power per log interval
of wave number is given by
\begineq
\label{eq:sparsessamplenoise}
\sigma(\Delta^2_\psi) = 
\left(1 + {N P_\psi \over 4 \Theta^2 \av{\gamma_1^2} }\right)
{4 \av{\hat \gamma_1^2} \omega \Theta \over N \sqrt{\pi d\ln \omega}}
\endeq
(if there is insufficient rejection of high frequency fluctuations
one would need to estimate and then subtract an extra constant term from
(\ref{eq:Ppsisparse}), and there would also be an additional
contribution to (\ref{eq:sparsessamplenoise}) from the fluctuations
in the aliased pwer.)
The two terms in (\ref{eq:sparsessamplenoise}) have opposite
dependence on the overall
survey size $\Theta$, and the optimum choice (assuming
a fixed instrument and integration time) is when these two
contributions are equal: $\Theta_\opt = \sqrt{N P_\psi / 4 \av{\gamma_1^2} }$.
For high redshift sources, and for large $N$,
this would dictate a very sparse
sampling indeed: $f = 4 \av{\gamma_1^2} / (n P_\psi) \sim 10^{-2}$ for 
$n \sim 2 \times 10^5$ per square degree and $Z_s \sim 3$,
but if one also wants to use the same survey
to measure the shear
for brighter and nearer objects one should adopt a correspondingly
less sparse strategy.  As a specific example, if one were to 
spread ones fields over a square say 3 times wider than the filled
survey then the sampling uncertainty would decrease by a factor
3 while the `shot-noise' term would increase by a factor 3.  For $Z_s \simeq 3$
the former would still dominate, at least on large angular scales, and
the resulting uncertainty is shown
in figure  \ref{fig:ppsiplot4} as the error-bars attached to the
EdS spectrum.  As one can see, the sparse sample has considerable
improved precision and one also has better coverage of the
behaviour of $P_\psi(\omega)$ around the peak as the fundamental
frequency is now reduced by a factor 3 also.  To design a
truly optimal survey for large-scale structure one really needs to
know the true level of small scale power, so it would be prudent
to first perform a smaller scale filled survey to establish this empirically.

\subsubsection{LSS and Amplification}

To close this discussion of large-scale structure, 
we consider what extra information can be gleaned
from estimates of the surface density derived from
the amplification.  In general, we expect the rms convergence
to be equal to the rms shear, and the convergence can be measured
directly as it causes a modulation of the counts of faint
galaxies $\Delta n / n = -2(1 - \alpha) \kappa$, where 
$\alpha = d \ln N(>l) / d \ln l = 2.5 d \log N / d {\rm mag}$ 
is the logarithmic slope of the counts and
the factor $-2(1 - \alpha)$ is known as the `amplification bias factor'.
At the magnitudes relevant for weak lensing observations 
$d \log N / d {\rm mag} \simeq 0.3$ (we will discuss the
dependence of $\alpha$ on waveband and colour presently)
so large-scale structure will give rise to apparent clustering of galaxies on
the sky with amplitude $\Delta n / n \simeq -0.5 \kappa$.  

Now for a power-law spectrum of mass fluctuations
in an Einstein - de Sitter cosmology the variance in $\kappa$
grows as $\zbar^{1-n}$, where $\zbar$ is the mean comoving distance to the
background galaxies (here $z = 1 - (1 + Z)^{-1/2}$).  On the other hand, 
if galaxies are unbiased tracers of the mass,
then $w(\theta)$ from the intrinsic spatial
clustering decreases as $(1 - \zbar)^{4}\zbar^{-(3+n)}$ (K92),
so, in this model, the relative importance of the lensing induced clustering
grows rapidly with increasing source redshift, and there
is a crossover at $Z_s \sim 2$ beyond which the induced clustering
comes to dominate (K92, Villumsen, 1996a).  
As discussed above though, a mass-traces-light normalisation seems
untenable for a high $\Omega_m$ universe and in either a low $\Omega_m$ 
universe, or a biased $\Omega_m = 1$ universe, the expected induced
clustering is sub-dominant.  In any case, it is not entirely clear how
one can separate the two effects observationally (without accurate
redshift information).

Another possibility is to 
cross correlate the surface density
$\kappa$ derived from lensing of faint galaxies with the surface brightness
for a sample of brighter objects (chosen so that their
$n(z)$ peaks where the $g(z)$ for the fainter background galaxies peaks).
The procedure for doing this with $\kappa$ derived from the shear
was described in K92, but this can equally well be done with $\kappa$ derived from
the amplification.  There is a slight complication here in that
the predicted effect is an anti-correlation of background and
foreground galaxy counts (as the bias factor is negative), and this will
be diluted (and perhaps overwhelmed) by contamination of the 
`background' sample by faint foreground galaxies, so an accurate
knowledge of the luminosity function is needed to properly account for this. 

For realistically normalised models we find rms $\kappa \sim 1$\% on degree
scales (for $Z_s \sim 3$).  For a shear based surface
density estimate $\kappa_\gamma$ the
uncertainty is $\sigma(\kappa_\gamma) \simeq \sqrt{\av{\gamma_1^2} / N}
\simeq 0.4 / \sqrt{N}$ and is much less than the expected signal for
a survey like that discussed above.  The  uncertainty
in the amplification derived surface density is
$\sigma(\kappa_A) = 0.5 (1 - \alpha)^{-1} \sqrt{(1 + n P_g) / N}
\simeq 2 \sqrt{(1 + n P_g) / N}$ 
where $P_g$ is the power spectrum of galaxy clustering, which can be the
angular power spectrum, in which case $n$ should be the density
of sources on the sky, or $P_g$ can be the spatial power spectrum,
in which case $n$ should be the three-dimensional number density,
the result being the same.
The ratio of the statistical uncertainty for the two techniques is
\begineq
{\sigma(\kappa_A) \over \sigma(\kappa_\gamma)} = 2 \sqrt{{1 + n P_g \over 
\av{\gamma_1^2} }} \simeq 5 \sqrt{1 + nP_g}
\endeq
The factor $1 + n P_g$ here (sometimes referred to as $\sim (1 + 4 \pi n J_3)$)
is an effective clustering multiplicity.  For a distribution of
randomly placed clusters,
it is just the number of galaxies per cluster,
and quite generally, the factor $nP_g$ tells us by how much the variance
of counts of galaxies exceeds that for a poisson distribution.  

For a power law angular correlation function 
$w(\theta) = w_0 (\theta / \theta_0)^{-0.8}$, as seems to be a reasonable
fit to the data, $n P \simeq 0.8 (k \theta_0)^{-1.2} 2 \pi n w_0 \theta_0^2$.
There are now a number of empirical estimates of $w(\theta)$ for
faint galaxies which we can use to estimate $2 \pi n w_0 \theta_0^2$ and
some recent estimates (with $\theta_0 = 1'$) are:
(1.52, Couch, Jurcevic and Boyle, 1993;
1.1  (B), 1.6 (R), Roche \etal, 1993;
1.45, Pritchet and Infante, 1992;
1.74, Efstathiou \etal, 1991;
3.8, Villumsen, Freudling and da Costa, 1996;
1.59, Brainerd, Smail and Mould, 1995).  
While these estimates span quite a range of magnitude limits
(the number density varying from $\sim  10$ per square arcmin
for the brighter surveys to $\sim 400$ for the Hubble Deep Field)
the estimates of $2 \pi n w_0 \theta_0^2$ are very stable
and indicate $nP \sim 2 (\omega 1')^{-1.2}$. 
These results are discouraging in the extreme. 
Even on sub-arcmin scales where the
clustering becomes negligible, the noise in the surface density inferred
from amplification is already about 5 times that for the shear based
$\kappa$ estimates (so the extra information contained in the
amplification is meagre), and on larger scales the situation
rapidly deteriorates.  
For cluster lensing (where one is probing structure on a
scale $\sim 5'-10'$) and where it might have been hoped that
amplification might resolve the `mass-sheet
degeneracy' problem, clustering has already inflated the
uncertainty by a factor 2 or so, so a cluster like A1689, which can be
detected at the 10-sigma level in the shear, would only be detectable at the 
$\sim 1 \sigma$ level (i.e.~the diminution of the counts
from lensing is about equal to the rms fluctuation expected from galaxy
clustering alone).  Worse still, for the degree scales of interest
here $\omega \sim 100 \rad^{-1} \sim 0.03 \arcmin^{-1}$
and we expect $n P$ to have grown to around $120$, so the clustering
fluctuations exceed the poisson fluctuations by about an order of magnitude.
We reach a very similar conclusion if we use Peacock's linearised
estimate of $P$; which is not surprising as this power spectrum model is
designed to fit the empirical data.  Thus, on degree scales, one would expect
$\sigma(\kappa_A) \sim 50 \sigma(\kappa_\gamma)$; so the extra information
in the
amplification based $\kappa$ estimate is negligible and, for surveys of the scale
envisaged here, one would expect at best a marginal detection
of the effect. The relatively high precision allowed
by the shear based surface density estimate relies on the assumption that
the intrinsic shapes of galaxies are uncorrelated, and could
be compromised if in fact there are strong intrinsic alignments
of galaxies
on supercluster scales.  There have been
a number of attempts to detect correlated orientations of galaxies
in superclusters --- with the hope of distinguishing between
`top-down' and `bottom-up' structure formation scenarios --- but no
convincing positive detections have been obtained
(see Djorgovski, 1986 for a review), and there are no indications
from weak lensing observations
(from e.g.~the rotation test) for any intrinsic alignments
at problematic levels.   

The counts slope is  somewhat waveband
dependent (the counts in $I$ are slightly flatter, and those in $V$
somewhat steeper than we have assumed).  This might suggest that one should use
the redder passband, and it has been suggested
(Broadhurst, 1996) that one would do better by selecting only red galaxies
as they have an even  shallower counts slope and hence a larger (negative)
amplification bias. However, this does not help here as it now appears
that faint red galaxies lie at relatively low redshift compared
to their bluer cousins (Luppino and Kaiser, 1996)
and this outweighs the gain from the flatter slope.
It is somewhat unfortunate, though not entirely coincidental, 
that the most distant objects 
(the faint and blue galaxies) which suffer the greatest amplification
have a very small amplification bias.

\section{Cluster Lensing}
\label{sec:clusterlensing}

We now consider lensing by an individual cluster which
is assumed to dominate over the effect of foreground and background
clutter.  From (\ref{eq:distortiontensor}), and specialising to
deflections occurring in a single plane, we can write
\begineq
\kappa = {\psi_{xx} + \psi_{yy} \over 2}
= {\Sigma \over \Sigma_\crit}
\endeq
where $\Sigma$ is the surface mass density and
$\Sigma_\crit$ is the critical surface density:
\begineq
\label{eq:sigmacrit}
\Sigma_\crit = {H_0 \sqrt{1 - \Omega_0} (1 + Z_l)
 \sinh \z_s \over 4 \pi G \sinh \z_l
\sinh(\z_s - \z_l)}
\endeq
which we have plotted against source redshift for a variety of lens
redshifts in  figure \ref{fig:sigmacritplot}.

For low $Z_l$ the impact of cosmology is very weak,
regardless of source redshift, but for $Z_s \sim 1$ or higher
we find a significant reduction in $\Sigma_\crit$ in  the
$\Lambda$ dominated model (due to the increased path length), and
an observer living in such a universe but using the
EdS $\Sigma_\crit$ would overestimate the mass of a cluster
at $Z \sim 1$ by almost a factor 2.

\beginfig
\plotone{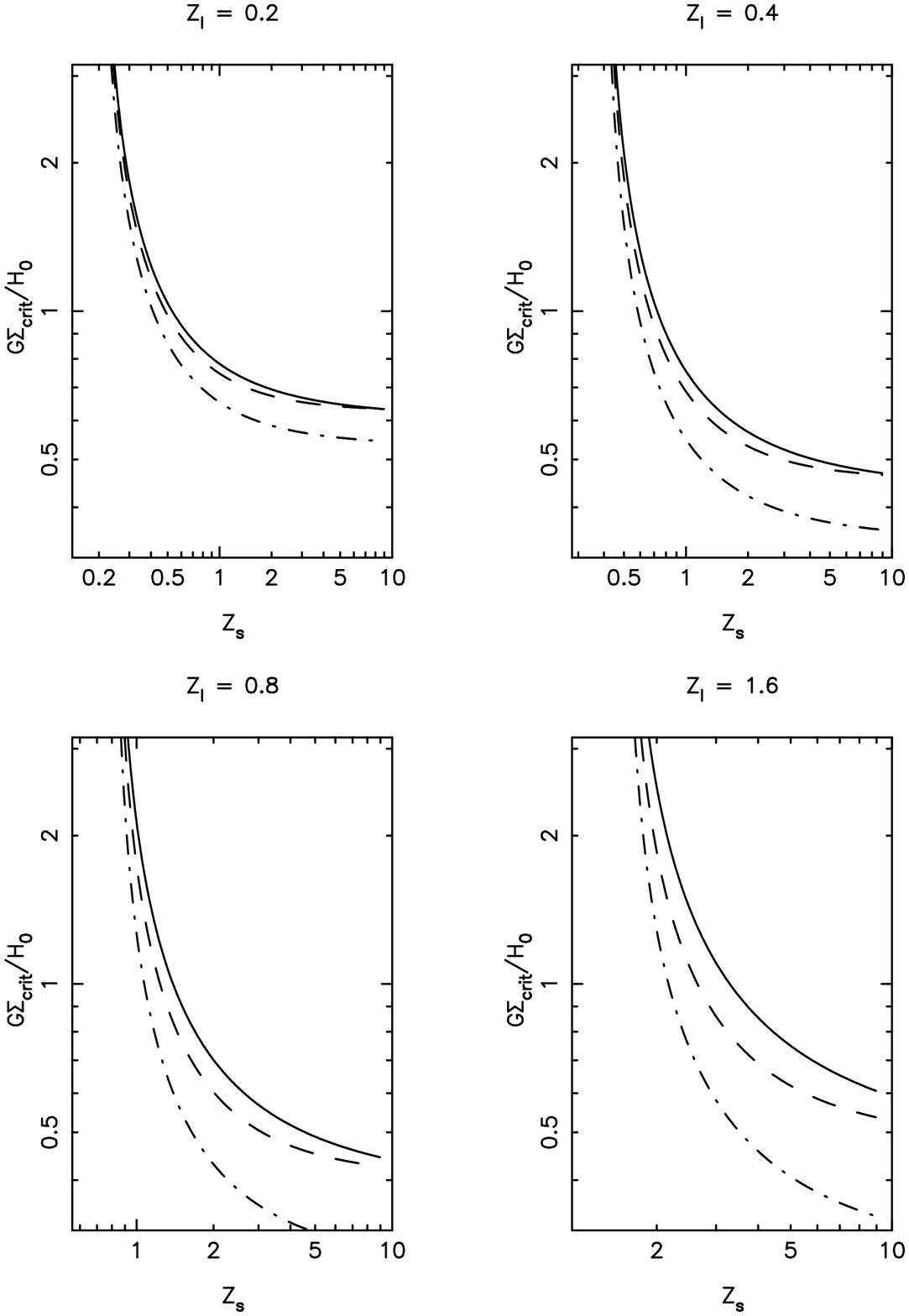}
\figcap{Critical surface density versus source redshift for
various lens redshifts and for our three illustrative
cosmological models.
\label{fig:sigmacritplot}}
\endfig

It is also interesting to compare the mass inferred from
lensing with that obtained from virial analysis
and/or X-ray temperature information.
Virial analysis gives
\begineq
\Sigma = {\alpha \sigma^2 \over G \theta a(\z_l) \sinh \z_l}
= {\alpha H_0 \sigma^2 \sqrt{1 - \Omega_0} (1 + Z_l) \over G \sinh \z_l}
\endeq
where $\alpha$ is some number of order unity which
accounts for the radial profile of the cluster;
velocity dispersion anisotropy; departures from sphericity;
departures from equilibrium; substructure; mass/light segregation
etc.~etc., so assuming that this can be done to sufficient
accuracy, we should find
\begineq
{\kappa \theta \over 4 \pi \alpha \sigma^2} = \beta(\z_l, \z_s) =
{\sinh(\z_s - \z_l) \over \sinh \z_s}
\endeq
This dimensionless ratio is an observable, and is
dependent on the cosmology. 
To see how useful this is we have plotted $\beta$ as
a function of $Z_l$, $Z_s$ for the three fiducial cosmological
models.  It is clear that the effect of varying the
cosmological parameters on this quantity is very weak.

\beginfig
\plotone{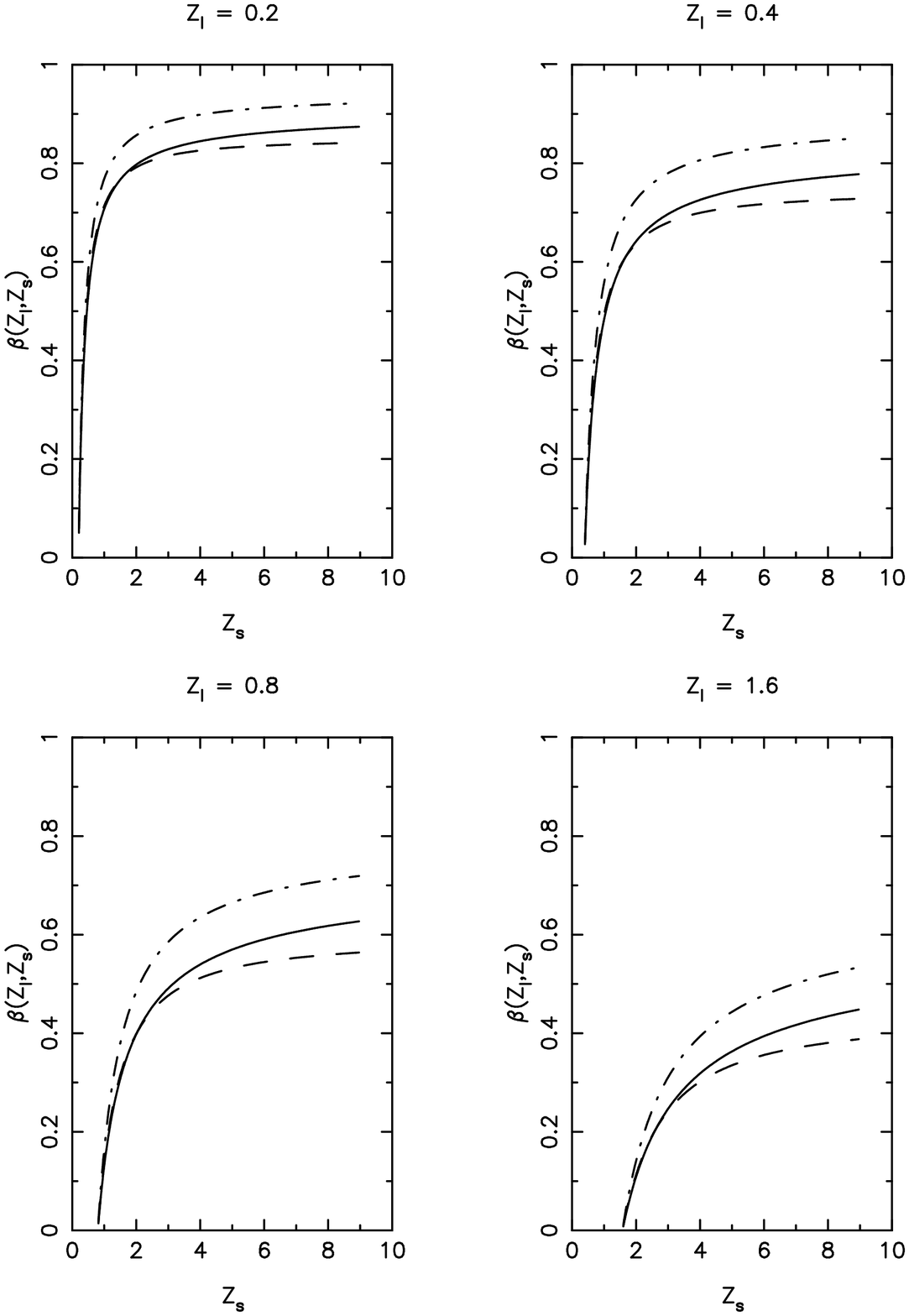}
\figcap{Relative distortion strength parameter $\beta$ versus
source redshift for various lens redshifts and for various values of
$\Omega_0$.
\label{fig:betaplot}}
\endfig

\newpage
\section{Summary}

The main new analytic result of this paper is the expression
(\ref{eq:Ppsidef}) which gives the angular power spectrum of the
distortion in terms of the 3-dimensional power spectrum of
potential fluctuations (a somewhat more useful form
for predicting $P_\psi(\omega)$ from models for the
density contrast power spectrum $P_\delta$ is given in
(\ref{eq:PpsifromPdelta})).  We have used this to compute
the distortion power for a number of illustrative models.

We have shown that for a sharply peaked 3-dimensional
spectrum, the scale on which the distortion appears is
cosmology dependent, being up to a factor $\sim 2$ smaller
in low density models. This is reflected in the model
predictions shown in figure \ref{fig:ppsiplot4} based on
Peacock's models for $P(k)$ which has a knee-like feature at
$\lambda \sim 100h^{-1}$Mpc, and consequently the location
of the corresponding
peak in $P_\psi(\omega)$ is comsology dependent.  
A more powerful cosmological discriminator is 
the growth of the distortion
with redshift, this being stronger in low matter density models
in general and in $\Lambda$-dominated models in particular.
This requires that one have at least approximate estimates
of the redshift of the faint galaxies as a function of flux, 
but this should be
feasible using approximate
redshift estimates by fitting
multicolour photometry to template spectra; 
Loh and Spillar, 1986;
Conolly, \etal, 1995;
Sawicki \etal, 1996)
This cosmological test does not require
any external normalisation of the power spectrum, but does require that
we should be able to measure the distortion with sufficient
precision at both high and low redshift.

One can also ask whether one can
hope to pin down the cosmology by making use of
external normalisation. This is certainly possible in
principle, but it is not yet entirely clear what is the appropriate
normalisation.  
Bernardeau \etal\ normalised to a fixed amplitude for
the density contrast and consequently found a
very strong dependence of the predicted shear on  the matter
density: $P_\psi \propto \Omega_m^{1.5}$ or thereabouts.
We have argued that this normalisation is unrealistic: if
instead we normalise to
galaxy clustering with a scale invariant bias and mass to light ratio
fixed by small scale cosmic virial theorem analysis then
we reach the opposite conclusion: low matter density models
in general and in $\Lambda$-dominated models in particular
then predict much stronger distortion at high redshift (the distortion
being cosmology independent for low redshift).  If on the other hand,
one were to normalise to some value for the amplitude of large-scale
bulk flows (still unfortunately a rather uncertain quantity) then 
$P_\psi \propto \Omega_m^{0.8}$ for very low source redshift, 
but for $Z_s \sim 1-3$ and
for realistic spectral indices  $n$ around $-1$ to $-2$ the predicted
distortion is only very weakly cosmology dependent.  A very similar
result is obtained if one normalises to cluster abundances.
This weak dependence on cosmology was also apparent when we computed
the distortion power for Peacock's fit to galaxy clustering
data, where all three illustrative models agree in distortion
power to within a factor 2. The high density model assumed a rather
mild bias $b = 1.6$, and for stronger bias the difference between the
models would be even less.

With a realistic normalisation we predict rms shear at the $\sim 1$\%
level at degree scales for sources at $Z_s \sim 3$, which should
be detectable at the $\sim 100$-sigma level with a survey
covering $\sim 10$ square degrees (which would contain $\sim  2 \times 10^6$
galaxies).  We found, however, that for a filled survey
of this size the sampling uncertainty
would much larger than the measurement noise, particularly at the
largest scales which in some ways are the most interesting.  
For some applications the sampling uncertainty is irrelevant,
but for the tests described above it is a serious handicap.
The sampling noise can be reduced considerably by adopting a sparse
sampling strategy (at some small cost in increased measurement
noise).  An important constraint on the design of such sparse surveys is aliasing of
power from small scales. While in principle this can be measured and subtracted
to obtain a fair estimate of the true large-scale power, it is still
an unwanted complication and, if the aliased power is dominant,
the precision will be compromised.
Peacock's empirically based models for the linear power spectrum
predict very low power at high frequencies and would
favour very sparse sampling (for deep surveys at least), but
safest approach is to measure the high frequency power directly
with a filled survey, and use this to determine the optimal sampling rate.

Finally, we considered the impact of cosmology on mass estimates
for individual clusters. We found that the critical surface density
was very similar in the matter dominated models, but is considerably
lower for high redshift lenses in a $\Lambda$-dominated model.
This is relevant to the result of Luppino and Kaiser (1996)
who found a strong shear signal for the cluster ms1054 at 
$Z_l = 0.83$.  In a $\Lambda$ dominated model, the mass
for this cluster would be reduced by about 40\%.
We also explored how the ratio of the lensing virial mass
(or mass inferred from X-rays) depends on cosmology, but found this
to be a very weak effect.

\end{document}